\documentclass[11pt]{article}
\pdfoutput=1 
\usepackage{jheppub}

\usepackage{amsmath}    % need for subequations
\usepackage{graphicx}   % need for figures
\usepackage{verbatim}   % useful for program listings
\usepackage{color}      % use if color is used in text
\usepackage{subfigure}  % use for side-by-side figures
\usepackage{hyperref}   % use for hypertext links
\usepackage{multicol}
\usepackage{amsfonts,amsbsy}
\usepackage{blkarray}

\definecolor{dgreen}{rgb}{0.0,0.6,0.0}

\def\beq{\begin{equation}\begin{aligned}}
\def\eeq{\end{aligned}\end{equation}}
\def\OO{\mathcal{O}}

\newcommand{\DM}{\text{DM}}
\newcommand{\DS}{\text{DS}}
\newcommand{\SM}{\text{SM}}
\newcommand{\NR}{\text{NR}}
\newcommand{\MP}{M_{\text{Pl}}}

\begin{document}

\title{Flooded Dark Matter and S Level Rise}
\date{\today}
\author[1]{Lisa Randall,} 
\author[1]{Jakub Scholtz,}
\author[2]{and James Unwin}
\affiliation[1]{Department of Physics, Harvard University, Cambridge, MA 02138, USA}
\affiliation[2]{Department of Physics,  University of Illinois at Chicago, Chicago, IL 60607, USA}

\abstract{
Most dark matter models set the dark matter relic density by some interaction with Standard Model particles. Such models generally assume the existence of Standard Model particles early on, with the dark matter relic density a later  consequence of those interactions. Perhaps a more compelling assumption is that dark matter is not part of the Standard Model sector and a population of dark matter too is generated at the end of inflation. This democratic assumption about initial conditions does not necessarily provide a natural value for the dark matter relic density, and furthermore superficially leads to too much entropy in the dark sector relative to ordinary matter. We address the latter issue by the late decay of heavy particles produced at early times, thereby associating the dark matter relic density with the lifetime of a long-lived state. This paper investigates what it would take for this scenario to be compatible with observations in what we call Flooded Dark Matter (FDM) models and discusses several interesting consequences.  One is that dark matter can be very light and furthermore,  light dark matter is in some sense the most natural scenario in FDM as it is compatible with  larger couplings of the decaying particle. A related consequence is that the decay of the field with the smallest coupling and hence the longest lifetime dominates the entropy and possibly the matter content of the Universe, a principle we refer to as ``Maximum Baroqueness''.  We also demonstrate that the dark sector should be colder than the ordinary sector, relaxing the most stringent free-streaming constraints on light dark matter candidates. We will discuss the potential implications for the core-cusp problem in a follow-up paper. The FDM framework will furthermore have  interesting baryogenesis implications. One possibility is that dark matter is like the baryon asymmetry and both are simultaneously diluted by a late entropy dump. Alternatively, FDM is compatible with an elegant non-thermal leptogenesis implementation in which decays of a heavy right-handed neutrino lead to late time reheating of the Standard Model degrees of freedom and  provide suitable conditions for creation of a lepton asymmetry.}

\setcounter{tocdepth}{2}

\maketitle

%%%%%%%%%%%%%%%%%%%%%%%%%%%%%%%%%%%%%%%%%%%%
%%%%%%%%%%%%%%%%%%%%%%%%%%%%%%%%%%%%%%%%%%%%

%%%%%%%
\vspace{-5mm}
%%%%%%%

\section{Introduction}

Most dark matter models set the dark matter relic density by some interaction with Standard Model particles. Such models generally assume the existence of Standard Model particles early on -- say immediately after inflation -- with the dark matter density set by those interactions later on.  Perhaps a more compelling assumption is that dark matter is not part of the Standard Model sector and a population of dark matter is generated at the end of inflation too. In this paper we take an agnostic point of view about the initial conditions of the Universe and dark matter's interactions with the Standard Model and assume that the inflaton decays democratically to the Standard Model and the dark sectors (if more than one). We call the dark matter and Standard Model particles produced directly through the inflaton decay the primordial matter, which includes a contribution to the dark matter relic density. We then ask what is required to reconcile this more compelling assumption with the currently observed Universe under the further assumption that   the comoving number density of dark matter remains constant (aside from a possible independent entropy dump into the dark sector too), i.e.~dark matter does not undergo thermal freeze-out, decay, or freeze-in production. In a future publication, we will show that with different couplings between the visible and dark sectors, our framework can also produce initial conditions compatible with these possibilities.

%%%%%%%%%%%%%%%%%%%%%%%%%%%%%%%%%%%%%%%
 \begin{figure}[t!]
\begin{center}
\includegraphics[height=55mm]{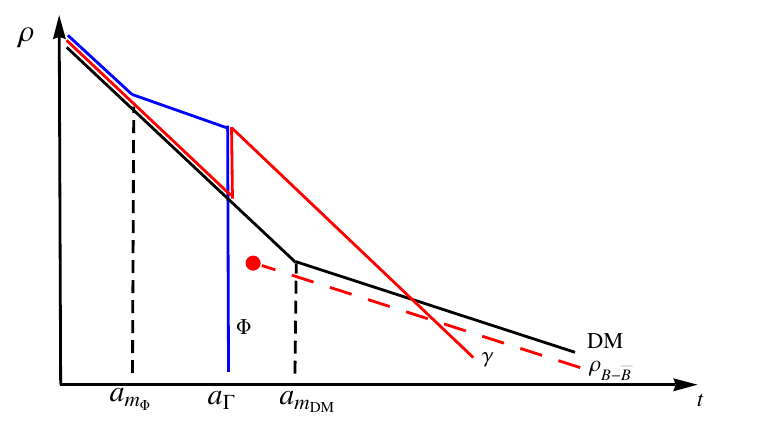}
\caption{Schematic plot of the time evolution of the energy densities, showing the dark matter (DM), Standard Model photon bath $\gamma$, and heavy states $\Phi$, for comparable initial densities $R^{(0)}\equiv\frac{\rho_\DM}{\rho_\Phi}\simeq1$. Also shown is the evolution of the net baryon number $\rho_{B-\overline{B}}\sim m_B n_B$, which for definitiveness we assume here is generated at some point following $\Phi$ decays and mark by the red dot.
\label{Fig1}}
\end{center}
\end{figure}
%%%%%%%%%%%%%%%%%%%%%%%%%%%%%%%%%%%%%%%

Given the above assumptions and  initial conditions, without a large injection of entropy into the Standard Model, dark matter would typically carry too much entropy. We therefore assume  the late decay of a heavy field $\Phi$ produced at early times adds entropy to the Standard Model at late times, thereby associating the dark matter relic density with the lifetime of a long-lived particle. By definition relativistic particles, including any light primordial states, redshift as radiation, $\rho \propto a^{-4}$, while any heavy species $\Phi$ becomes nonrelativistic and evolves like matter, $\rho \propto a^{-3}$.  As a result the contribution of the primordial states to the total energy density will rapidly diminish. The entropy resulting from the late decay can therefore flood the entropy of the universe, adequately diluting the primordial dark matter contribution. Hence, we call this scenario {\em Flooded Dark Matter} (FDM).

For simplicity, and because it embodies much of the key physics, we will initially consider only one such heavy state $\Phi$. After the decay of $\Phi$ into Standard Model particles, the subsequent evolution of dark matter and the  Standard Model will  behave in accordance with standard cosmology. The relative evolution of the densities sets the abundance of dark matter and photons, as in Figure \ref{Fig1}. Assuming very heavy $\Phi$, the period for which the energy densities undergo different evolution is determined by the lifetime of $\Phi$. The entropy injection due to $\Phi$ decays reheats the visible sector so that the Standard Model dominates the entropy. This also explains why radiation contributions from other sectors (``dark radiation'') should not influence cosmological observables, as these are diluted by the entropy dump too. If there are several heavy states, the final species to decay typically contributes the most energy and entropy, as energy injected in prior decays is also diluted relative to the energy in the remaining nonrelativistic states. Other authors \cite{moduli,Moroi:1999zb} have considered the cosmological implications of entropy dumps by scalar fields primarily in the context of moduli decay. In contrast our $\Phi$ is in general not a string modulus (or even a scalar), its lifetime is much shorter and it generally decays preferentially into a particular sub-sector.\footnote{Indeed, it is likely difficult to realise FDM with $\Phi$ identified as a string modulus since, even utilising sequestering effects, it is challenging to avoid sizeable branching ratios to all sectors in the moduli decays \cite{Moroi:1999zb}.} In the spirit of democracy we will ultimately assume at least two species: one that decays into dark matter $\Phi_\DM$ and $\Phi_\SM$, which decays into Standard Model particles.

Our analysis shows how to make the inflaton democratic decay consistent with both the relative entropy and energy of the dark and ordinary matter sectors.  An interesting consequence  of this scenario is that FDM can be very light in this scenario and we will explain how light dark matter appears naturally in our framework. We will show that the dark sectors are typically colder than the visible sector, which relaxes the stringent free-streaming constraints on light dark matter candidates. A consequence of this possibility is that Fermi-blocking with light dark matter can conceivably solve the core-cusp problem; we will address this in a dedicated paper \cite{LJJ}. With low dark matter mass, we find that the entropy in the dark sector lies between that of baryons and photons, but is not tied to either. 

A natural question in this context is how to reconcile this with the baryon asymmetry. Our framework introduces a few interesting possibilities:

\begin{enumerate}

\item A particle asymmetry comes either from inflaton decays or from dynamics in the early Universe and is initially present (as in asymmetric dark matter models)  in both the  visible  and dark matter sector (see e.g.~\cite{inflaton-baryo}). The asymmetry in this case would be carried either by $\Phi$ or by the primordial states -- the difference from asymmetric dark matter  simply being the later dilution. This might in some sense be the most compelling scenario in our context because baryons and dark matter, which both carry less entropy than photons, are treated on the same footing and diluted by the late decay of $\Phi$. 

\item CP violating decays of $\Phi$ to the Standard Model generate an asymmetry in $B$ or $L$,  similar to e.g.~\cite{Davidson:2008bu,leptog,other}.  This option leads to more model-dependent parameters and constraints but can have interesting implications. We will present a realization based on a conventional see-saw neutrino model in which the heavy right-handed fermion plays the role of $\Phi$, and the electron right-handed neutrino, which is most weakly coupled, is responsible for the Standard Model content.

\item The baryon asymmetry is generated by dynamics in the visible sector. This might occur after reheating due to $\Phi$ decays (e.g.~via electroweak baryogenesis \cite{Morrissey:2012db}) or prior to the decay of $\Phi$ in which case any asymmetry will be subsequently diluted. This scenario is not tied to our framework and thus we do not discuss it further.

\end{enumerate}

We emphasize our goal -- unlike most dark matter models -- is not to naturally explain the amount of energy remaining in the dark sector, although it is interesting that some models, like the neutrino model we present later, do work beautifully. We are simply exploring what it would require for a democratically decaying inflaton past to match onto the Universe we see today and furthermore explore several interesting consequences.

The paper is structured as follows: in Section~\ref{sec:framework} we derive an expression for the dark matter relic density in terms of the $\Phi$ decay rate. We subsequently examine the observational implications of FDM -- specifically the bounds from free streaming and contributions to the effective number of neutrino species. In Section~\ref{sec:baryogenesis} we investigate the different implementations of baryogenesis as outlined above. In Section~\ref{sec:neutrino} we construct an elegant realization of this framework in which a heavy state $\Phi$ is identified with a right-handed neutrino. We conclude in Section~\ref{sec:conclusion}. Some further relevant details are given in the Appendices.
 
%%%%%%%%%%%%%%%%%%%%%%%%%%%%%%%%%%%%%%%%%%
%%%%%%%%%%%%%%%%%%%%%%%%%%%%%%%%%%%%%%%%%%%

\section{Flooded Dark Matter}
\label{sec:framework}

In this section we  give a description of the evolution of the Universe within the FDM framework and derive the relevant constraints on the parameter space. We first do this for a scenario with one heavy particle $\Phi$ in Section~\ref{ssec:onefield}, and later  generalize to a two particle scenario with $\Phi_\SM$ and $\Phi_\DM$ in Section~\ref{ssec:twofields}. We then investigate the constraints from dark matter free-streaming and the effective number of neutrino species $\Delta N_{\rm eff}$ in Section~\ref{ssec:freestreaming}. Finally, we collect all of the relevant constraints and relations in Section~\ref{ssec:parameterspace}.

%%%%%%%%%%%%%%%%%%%%%%%%%%%%%%%%%%%%%%%%%%%

\subsection{Standard Model Reheating from Late Decays}
\label{ssec:onefield}

We first consider a scenario with only a primordial dark matter contribution and make the simplest assumption of a single heavy scalar field that decays to the Standard Model, which in this section we will call $\Phi\equiv \Phi_\SM$.  The period for which the energy density of the dark matter redshifts relative to the energy density of $\Phi$  is controlled by the $\Phi$ lifetime so we first derive the $\Phi$ decay rate required to match the observed dark matter relic density. We denote the scale factor at which $\Phi$ becomes nonrelativistic ($T \simeq m_\Phi$) by $a = a_0$, and that at which $\Phi$ decays as $a = a_\Gamma$. The scale factor when dark matter becomes nonrelativistic will be denoted as $a = a_{\NR}$. Moreover, we define the ratio of energy densities of dark matter and $\Phi$ at different cosmological times as 
\begin{equation}
R^{(i)} \equiv R(a_i) \equiv \frac{\rho_\DM(a_i)}{\rho_\Phi(a_i)}  ~.
\label{RR}
\end{equation}
Assuming democratic inflaton decay $R^{(0)}\equiv R(a_0)\simeq1$. However, we shall leave $R^{(0)}$ as a free parameter as there are other motivated scenarios in which $R^{(0)}$ could deviate from unity, be near vanishing, or be related to the fundamental model parameters.\footnote{For instance, if only dark matter is reheated after inflation, freeze-in will subsequently generate a population $n_\Phi^{\rm FI}$ of $\Phi$ states. As a result $R^{(0)}\sim n_\Phi^{\rm FI}/s$ and $n_\Phi^{\rm FI}$ will depend on the couplings involved in the freeze-in portal operator (see e.g.~\cite{FI}).} Also, treating $R^{(0)}$ as a free parameter allows the possibility of different initial temperatures in the various sectors.

We might also wish to keep track of other primordial populations such as the Standard Model sector and any additional dark sectors. We mark the ratios of densities of these primordial populations to the density of $\Phi$ similarly
\begin{equation}
R^{(0)}_{\SM} \equiv \frac{\rho_{\SM}(a_0)}{\rho_\Phi(a_0)}; 
\hspace{15mm} R^{(0)}_{\DS} \equiv \frac{\rho_{\DS}(a_0)}{\rho_\Phi(a_0)}~.
\end{equation}
The evolution of  $\rho_{\rm tot}$ can be described in terms of the Hubble parameter as follows
\begin{equation}
H^2(a) =
\frac{\rho_{\rm tot}(a)}{ 3\MP^2} \simeq
\frac{g_{\Phi} \pi^2}{90}
\frac{m_\Phi^4}{\MP^2}\left[ \left(\frac{a_0}{a}\right)^3 + R^{(0)} \left(\frac{a_0}{a}\right)^4 + R^{(0)}_{\SM} \left(\frac{a_0}{a}\right)^4 + R^{(0)}_{\DS} \left(\frac{a_0}{a}\right)^4   \right]~,
\label{eq:fullhubble}
\end{equation}
where $M_{\rm Pl}$ is the reduced Planck mass. We denote by $g_{i}$ the number of relativistic degrees of freedom in a given state or sector $i$, scaled by $7/8$ for fermions.\footnote{We will neglect throughout the numerically small difference between the counting of fermionic degrees of freedom in number density and energy density.} Similar to our definition of $R^{(i)}$ we use a superscript to indicate the temperature at which a given quantity should to be evaluated, specifically, the value of $g$ at a given moment $a=a_i$ will be denoted $g^{(i)}$ and we use $g^{(\infty)}$ for its present-day value. 
Note that the first term in eq.~(\ref{eq:fullhubble}) corresponds to the energy density contribution from $\Phi$, while the second term corresponds to the contribution from the dark matter particle. The subsequent terms describe the contribution of the primordial Standard Model and additional dark sector populations. The decays of $\Phi$ become important when $3H(a_\Gamma) = \Gamma$ (for a  determination of the precise  numerical factor see Appendix \ref{AppendixA}). 

We assume that prior to $3H=\Gamma$ the state $\Phi$ dominates the energy density of the Universe and that at this point the dark matter is still relativistic. We show in Appendix \ref{AppendixB} that the latter assumption is unnecessary, but assume this now for clarity of exposition. Indeed, in the converse scenario, where the dark matter becomes nonrelativistic before $\Phi$ decays, one obtains the same result up to $\OO(1)$ numerical prefactors.

The energy density of Standard Model particles must exceed that of dark matter particles once $\Phi$ has decayed. As a result $\Phi$ has to dominate the total energy density at this point.\footnote{In the special case in which the primordial Standard Model population already dominates the energy density, we have no need for $\Phi$ and we therefore omit this case.} We drop all other energy contributions and set $3H=\Gamma$ in eq.~(\ref{eq:fullhubble}) to get the scale factor at time of $\Phi$ decay
 \beq
\left(\frac{a_0}{a_\Gamma}\right)^3=
\frac{10}{\pi^2}
\frac{\Gamma^2 \MP^2}{ g_\Phi m_\Phi^4}~.
\label{eq:aGamma} 
\eeq
Further, eq.~(\ref{eq:fullhubble}) determines the ratio of energy densities at the time of the $\Phi$ decay
\begin{equation}
R^{(\Gamma)}   = R^{(0)} \left(\frac{a_0}{a_\Gamma}\right)
= R^{(0)} \left[\frac{10}{\pi^2}
\frac{\Gamma^2 \MP^2}{g_\Phi m_\Phi^4}\right]^{1/3}~.
\label{eq:RGamma}
\end{equation}
Assuming the evolution of the Universe is adiabatic after $\Phi$ decays, the ratio of entropy densities does not change between $a_\Gamma$ and the present-day. Thus we can relate $R^{(\Gamma)}$ to the ratio of entropy densities of the reheated Standard Model population and dark matter today
\beq
R^{(\Gamma)} = \left(\frac{g_{\SM}^{(\Gamma)}}{g_{\DM}^{(\Gamma)}}\right)^{1/3} \left( \frac{s_\DM^{(\Gamma)}}{s_\SM^{(\Gamma)}}\right)^{4/3} 
= 
\left(\frac{g_{\SM}^{(\Gamma)}}{g_{\DM}^{(\Gamma)}}\right)^{1/3} 
 \left(\frac{s_\DM^{(\infty)}}{s_\SM^{(\infty)}}\right)^{4/3}~.
\label{eq:rgamma}
\eeq
Moreover the ratio of dark matter to Standard Model entropies can be expressed in terms of  observed quantities as follows
\beq
\frac{s_\DM^{(\infty)}}{s_\SM^{(\infty)}} = \frac{2\pi^4}{45 \zeta(3)} \Delta \frac{ n_\DM}{n_B} = \frac{2\pi^4}{45 \zeta(3)} \Delta \frac{ \Omega_\DM}{\Omega_B}\frac{m_B}{m_\DM},
\label{eq:sos}
\eeq
where $\Delta = n_B/s_\SM = 0.88 \times 10^{-10}$ and $m_B\approx0.938$ GeV is the proton mass.  Collecting eqs.~(\ref{eq:RGamma})~and~(\ref{eq:rgamma}) we obtain an expression for the decay rate of $\Phi$ required to match the observed relic density today 
\begin{equation}
 \Gamma = 
 \frac{\pi}{\sqrt{10}}
 \frac{m_\Phi^2}{M_{\rm Pl}}
  \left(\frac{s_\DM^{(\infty)}}{s_\SM^{(\infty)}}\right)^2
  \left(\frac{g_\SM^{(\Gamma)}}{g_\DM^{(\Gamma)}} \right)^{1/2}  \left(\frac{g_\SM^{(\Gamma)} }{\left(R^{(0)}\right)^3}\right)^{1/2}~.
\label{eq:Ga-0}
\end{equation}

\begin{figure}[t!]
\begin{center}
\includegraphics[height=60mm]{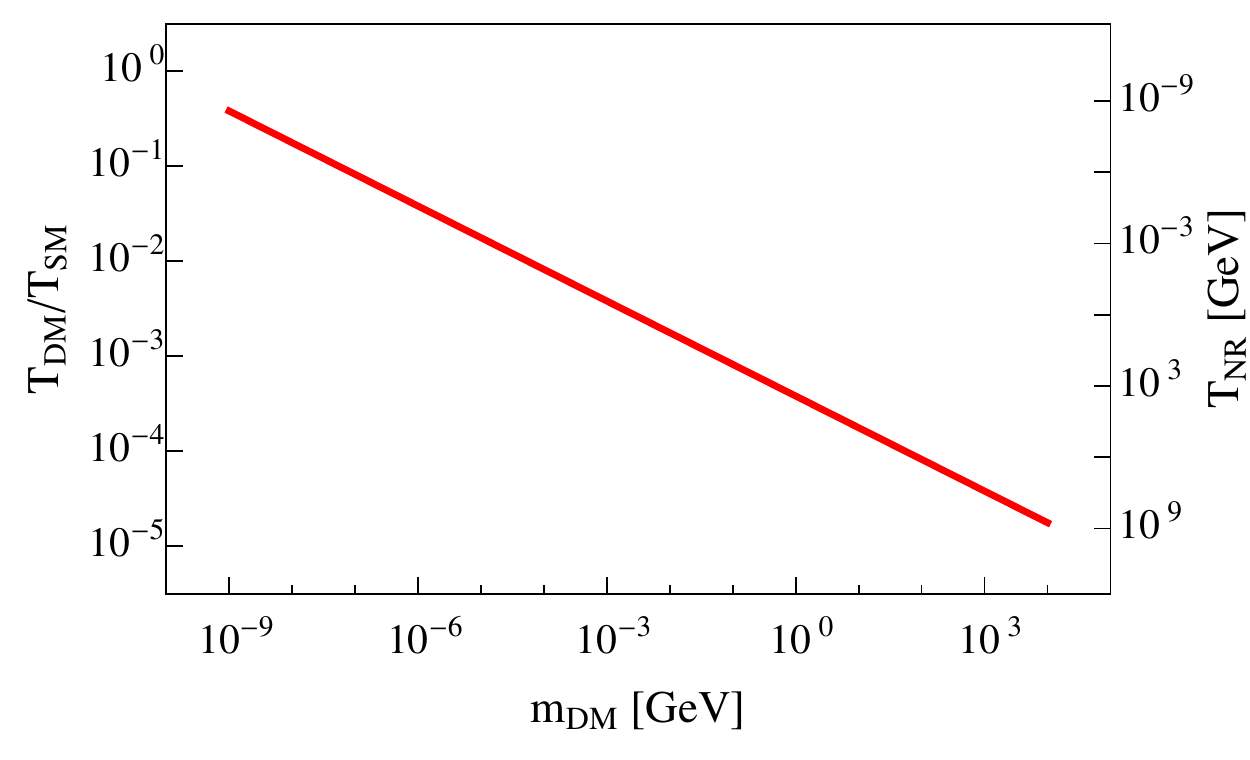}
\caption{Ratio of dark to visible sector temperatures $T_{\rm DM}/T_{\rm SM}$ as a function of $m_\DM$. The right-hand $y$-axis shows the temperature of the Standard Model $T_{\NR}$ when dark matter becomes nonrelativistic.
\label{Fig2}}
\end{center}
\end{figure}

Additionally, from inspection of eq.~(\ref{eq:rgamma})~and~(\ref{eq:sos}) we verify that $R^{(\Gamma)}\ll1$, as long as $m_\DM \gg 1$~eV, thus supporting our decision to drop the energy term associated with dark matter in eq.~(\ref{eq:fullhubble}). Furthermore, $R^{(\Gamma)}\ll1$ implies that the energy density of the dark matter is small compared to the Standard Model at this point.  This is different from thermal dark matter for which one might expect that the Standard Model and dark matter might have been in thermal equilibrium until around $T\sim m_\DM$, as a result the dark matter in FDM models can be significantly colder than expected for thermal dark matter. The ratio of the temperatures of the dark matter sector and Standard Model sector is simply
\beq
\begin{aligned}
\left.\frac{T_{\DM}}{T_{\SM}}\right|_\infty = \left(\frac{g_{\SM}^{(\Gamma)}}{g_{\DM}^{(\Gamma)}}R^{(\Gamma)} \right)^{1/4} \left(\frac{g_{\SM}^{(\infty)}}{g_{\SM}^{(\Gamma)}}\right)^{1/3} 
&= \left(\frac{g_{\SM}^{(\infty)}}{g_{\DM}^{(\Gamma)}}\frac{2\pi^4}{45 \zeta(3)} \Delta \frac{ \Omega_\SM}{\Omega_B}\frac{m_B}{m_\DM}\right)^{1/3}~.
\label{eq:tdmtsm}
\end{aligned}
\eeq
In Figure \ref{Fig2} we show $T_{\rm DM}/T_{\rm SM}$, as a function of the dark matter mass, which follows from eq.~(\ref{eq:tdmtsm}).
At decay $\rho_\Phi\simeq\rho_\SM$ and thus the Standard Model reheat temperature can be expressed
\beq
T_{\rm RH} \equiv T_{\rm SM}^{(\Gamma)}
\simeq
\left[\frac{30}{\pi^2 g_{\SM}^{(\Gamma)}}  \rho_{\SM}(a_\Gamma)\right]^{1/4}
\simeq   \left[\frac{10}{\pi^2 g_{\SM}^{(\Gamma)}} \right]^{1/4}
\sqrt{\Gamma M_{\rm Pl}} ~.
\label{eq:SMrh}
\eeq
Note that to satisfy the constraints from BBN it is required that $T_{\rm RH} \gtrsim 10$ MeV  \cite{Cyburt:2015mya,Sarkar:1995dd}. However, in many of  the baryogenesis scenarios we consider in detail later in this paper we shall rely on sphaleron processes and thus we enforce the stricter condition $T_{\rm RH} \gtrsim 100$ GeV.

Finally, we wish to determine the visible sector temperature at which dark matter becomes nonrelativistic. This occurs once the dark sector temperature drops to the dark matter mass threshold $T_\DM \simeq m_\DM$, at this point the Standard Model temperature is given by
\beq
\begin{aligned}
T_{\rm NR} \simeq m_\DM \left. \frac{T_\SM}{T_\DM}\right|_{\rm NR} & 
= m_\DM \left(\frac{g_{\SM}^{(\Gamma)}}{g_{\DM}^{(\Gamma)}} \frac{s_\DM^{(\infty)}}{s_\SM^{(\infty)}} \right)^{-1/3} \left(\frac{g_{\SM}^{(\rm NR)}}{g_{\SM}^{(\Gamma)}}\right)^{-1/3} 
\\
&= m_\DM \left(\frac{g_{\SM}^{(\rm NR)}}{g_{\DM}^{(\Gamma)}}\frac{2\pi^4}{45 \zeta(3)} \Delta \frac{ \Omega_\DM}{\Omega_B}\frac{m_B}{m_\DM}\right)^{-1/3}.
\end{aligned}
\label{eq:tnr}
\eeq
The weak dependence on the ratio of degrees of freedom can typically be neglected.

%%%%%%%%%%%%%%%%%%%%%%%%%%%%%%%%%%%%%%%%%%%

\subsection{Dark Matter from Late Decay}
\label{ssec:twofields}

We have so far considered only one heavy field $\Phi$ that decays into Standard Model particles. Now we consider the more general possibility that the inflaton decays into dark matter particles, Standard Model particles, and at least two heavy fields $\Phi_{\DM}$ and $\Phi_{\SM}$ associated with the dark matter and Standard Model sectors.\footnote{Notably, in the context of supersymmetric models, the lowest dimension flat directions of the superpotential are good  candidates for $\Phi_\DM$ and $\Phi_\SM$ as they are more likely to decay into one sector or the other if the two sectors involve distinct sets of gauge interactions.} 
We assume that the second  field $\Phi_\DM$ that decays primarily to dark matter, and that the field decaying to the Standard Model is longer-lived. Hence the Standard Model entropy will dominate over that of the dark matter, that redshifts before the second field decays. This matches continuously on to the single decaying field model of the previous section, which can be considered a limiting case when the decay to dark matter occurs before $\Phi_\DM$ becomes nonrelativistic. 

When dark matter is produced by late decay, there is less time for dark matter to redshift relative to ordinary matter. For this reason, the allowed parameter space is reduced. Here we assume that $\Phi_{\SM}$ and $\Phi_\DM$ are degenerate. This reduces the allowed parameter space. We demonstrate this for the case when $\Phi_\SM$ and $\Phi_\DM$ are degenerate, and generalize to the nondegenerate case in Appendix \ref{ApC}.
So we consider  $m_{\Phi_\DM}=m_{\Phi_\SM}=m_\Phi$, but with $\Gamma_{\DM}>\Gamma_{\SM}$, where we denote $\Gamma_{\Phi_i}\equiv\Gamma_{i}$.  We will restrict our attention to scenarios in which both states decay at temperatures below the $m_\Phi$ threshold. Assuming that at the mass thresholds the Universe is matter dominated by $\Phi$, this occurs for $H\sim m_{\Phi}^2/M_{\rm Pl}$. We define $a_0\equiv a(T=m_\Phi)$, and take the initial conditions
\beq
\rho_i(a_0)=R^{(0)}_{i} m_\Phi^4~,
\eeq
where $R^{(0)}_{i}$ accounts for the initial ratios (for $i=\Phi_\DM,\Phi_\SM,$ DM, SM) and we absorb the $g_i$ factors into the definition of $R^{(0)}_{i}$ here.

The energy densities are evolved to $H\simeq\Gamma_{\DM}$  to obtain
\beq
\rho_{i}(a_{\Gamma_{\DM}}) &=R^{(0)}_{i} m_\Phi^4\left(\frac{a_0}{a_{\Gamma_{\DM}}}\right)^3~,
\hspace{8mm}
 (i=\Phi_{\DM},\Phi_{\SM}) ~.
\eeq
Because the dark matter redshifts like radiation between the time of the first decay to the time of the second, and this era is matter dominated, 
it is easy to see that after the second field has decayed 
\beq
\frac{\rho_\DM(a_{\Gamma_{\SM}})}{\rho_\SM(a_{\Gamma_{\SM}})} 
&= 
 \frac{R^{(0)}_{\Phi_\DM}}{R^{(0)}_{\Phi_\SM}}
\left[ \frac{R^{(0)}_{\Phi_\DM}+R^{(0)}_{\Phi_\SM}}{R^{(0)}_{\Phi_\SM}}
\left(\frac{\Gamma_{\SM}}{\Gamma_{\DM}}\right)^2 \right]^{1/3}~,
\label{eq-n1}
\eeq
where we have accounted for the possibility of different initial densities stored in the two fields, which enters both directly and in the energy stored in matter during the interval between the two decays. Notice that this is the same form as eq.~(\ref{eq:RGamma}) above with the substitution $m_{\Phi}^2/M_{\rm Pl} \to \Gamma_{\DM} $ since the relative redshifting no longer starts right after $\Phi$ becomes nonrelativistic, but after $\Phi_{\DM}$ decays. From eq.~(\ref{eq:sos}) \& (\ref{eq-n1}) we find that for a given  $\Phi_\DM$ decay rate, the required $\Gamma_{\SM}$ to reproduce the observed dark matter relic density 
\beq
\Gamma_{\SM} =
 \Gamma_{\DM} 
  \left(\frac{ g_\SM^{(\Gamma)}}{g_\DM^{(\Gamma)}} \right)^{1/2}
  \left(\frac{s_\DM^{(\infty)}}{s_\SM^{(\infty)}}\right)^2 \left[\left( \frac{R^{(0)}_{\Phi_\SM}}{R^{(0)}_{\Phi_\DM}}\right)^{3} \frac{R^{(0)}_{\Phi_\SM}}{R^{(0)}_{\Phi_\DM}+R^{(0)}_{\Phi_\SM}} \right]^{1/2}
~.
\label{fi-1}
\eeq
This expression is the analog of eq.~(\ref{eq:Ga-0}) in the one-field FDM framework.

%%%%%%%%%%%%%%%%%%%%%%%%%%%%%%%%%%%%%%%
 \begin{figure}[t!]
\begin{center}
\includegraphics[height=75mm]{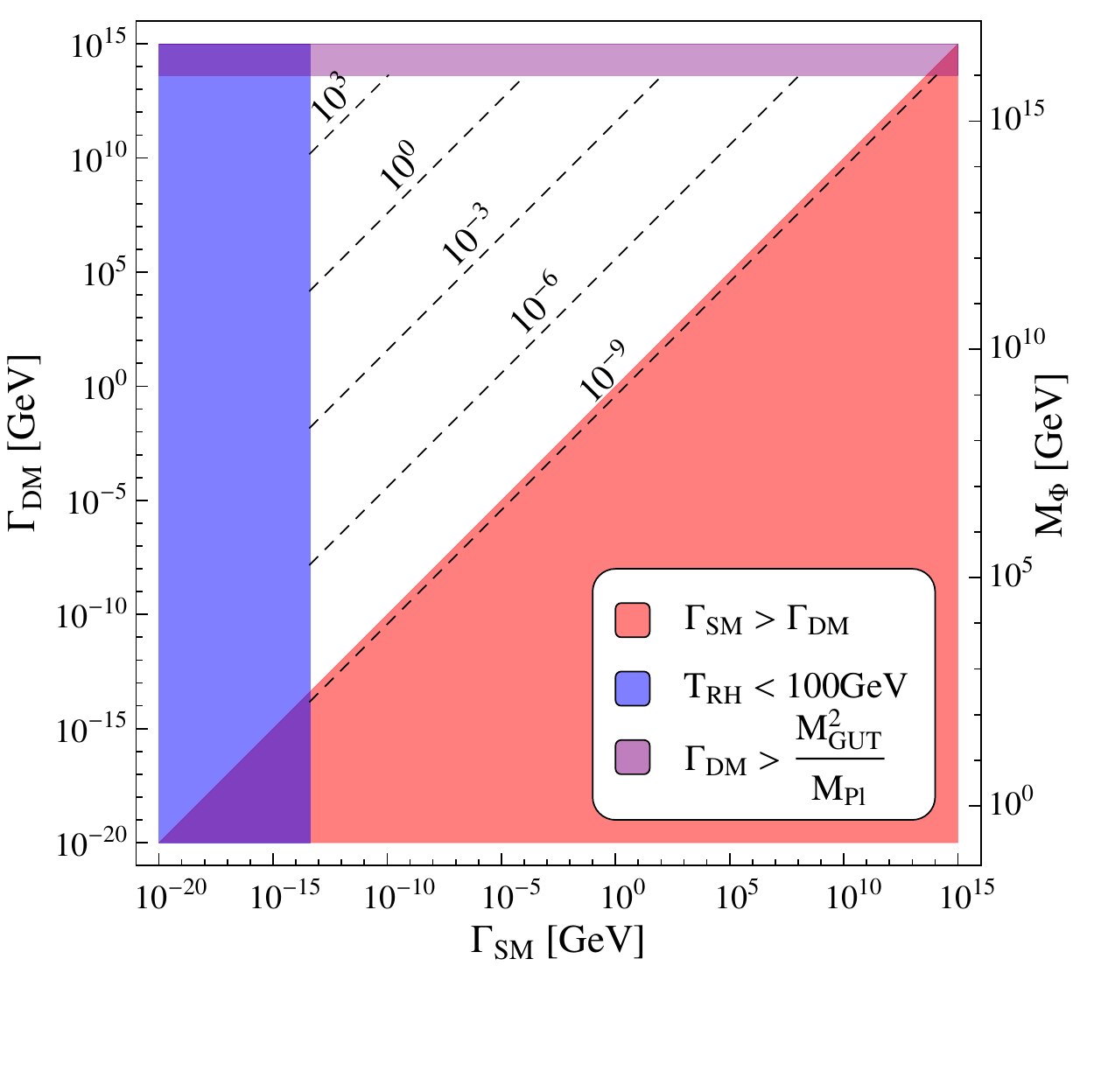}
~
\includegraphics[height=75mm]{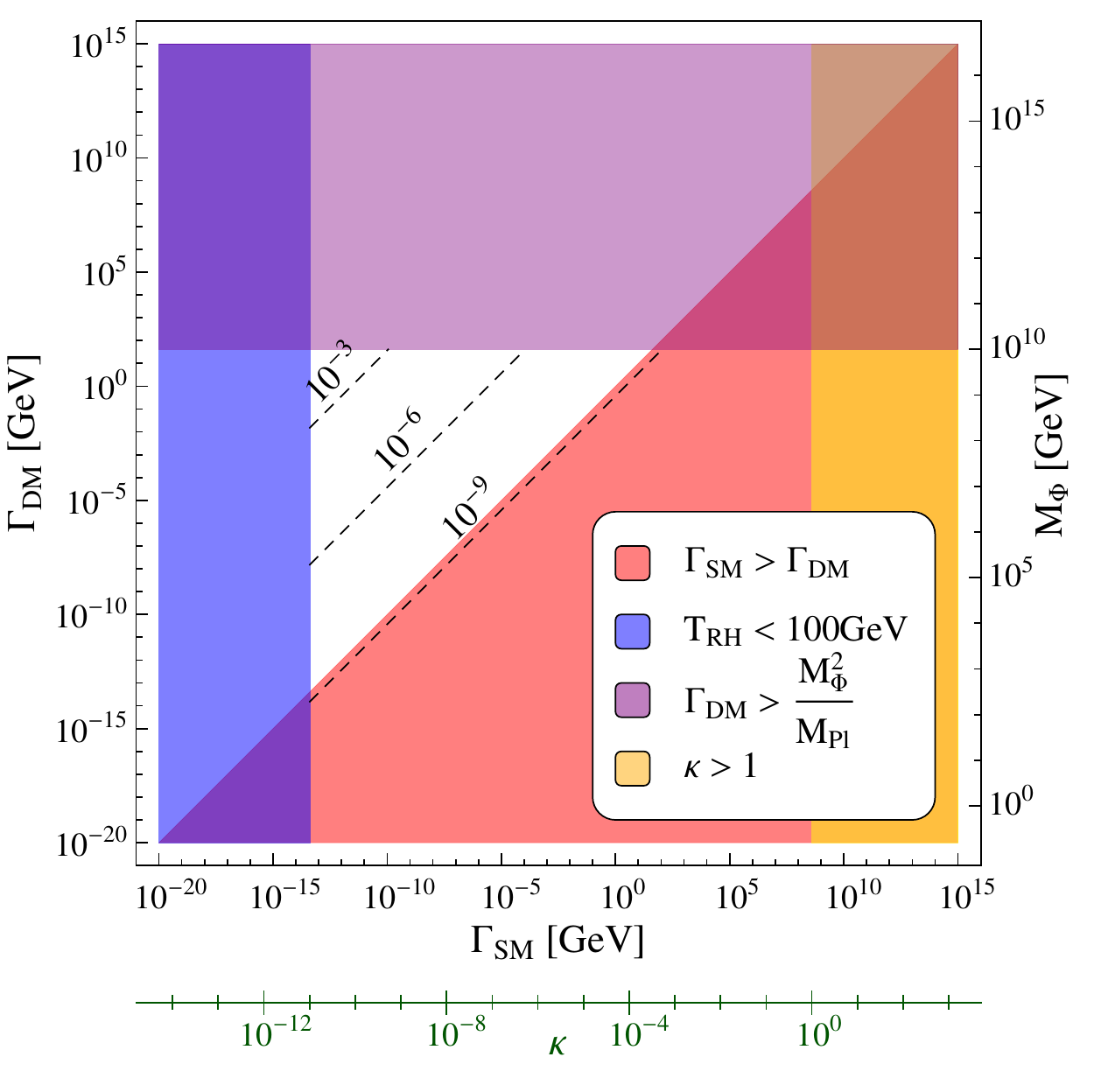}
\caption{
Left: The allowed parameter space in the $\Gamma_\SM$--$\Gamma_\DM$ plane, as constrained by the relic density of dark matter. The blue region indicates a reheat temperature below $100$ GeV. The purple indicates regions in which $\Phi_\DM$ decays earlier than the highest temperature the Universe can achieve in reasonable models of reheating: $\Gamma_\DM > H_{\rm IRH}$. The red indicates regions in which the $\Phi_\DM$ decays later than $\Phi_\SM$, although logically possible, we will see these regions are typically excluded. The dashed contours show the mass of dark matter $m_\DM$ (GeV) that is necessary to recover the observed relic energy density of dark matter. Finally, the vertical axis on the right shows the values of $m_\Phi$ in one field models that gives equivalent results to the choice of $\Gamma_\DM$ in two field framework.
Right: This plot is the same plot as the left panel with a particular choice of $m_\Phi = 10^{10}$ GeV. The dark green $x$-axis shows the values of $\kappa$ necessary to match $\Gamma_\SM = \kappa^2 m_\Phi/8\pi$. Notice that larger $\kappa$s correspond to small masses of dark matter. A perturbativity constraint $\kappa > 1$ is now marked in yellow.
\label{couplings}}
\end{center}
\end{figure}
%%%%%%%%%%%%%%%%%%%%%%%%%%%%%%%%%%%%%%%

The left panel of Figure~\ref{couplings} shows regions in the $\Gamma_\SM$--$\Gamma_\DM$ plane for which we recover the correct relic density of dark matter. We have incorporated the constraints that $\Phi_\SM$ is longer lived than $\Phi_\DM$, that the Standard Model reheat temperature is above $100$ GeV and that $\Phi_\DM$ decays after the initial reheat of the Universe. This latter condition is ensured be imposing that the decay rate of $\Phi_{\rm DM}$ should be larger than the Hubble rate at inflationary reheating: $\Gamma_\DM > H_{\rm IRH}$. As we have seen above, the two-field framework is equivalent to the one-field framework with mass $m_{\Phi}^2  = M_{\rm Pl} \Gamma_{\DM}$. We show this equivalent mass on the right-hand axis of each plot to facilitate the comparison. The right panel shows the consequences of choosing $m_\Phi = 10^{10}$ GeV. Once the common mass is chosen we can determine the coupling $\kappa$, defined as $\Gamma_\SM = \kappa^2 m_\Phi / 8\pi$ which is shown on the dark green $x$-axis below the plot. The dashed contours indicate $m_\DM$. Notice that the largest $\kappa$ require a small dark matter mass.

%%%%%%%%%%%%%%%%%%%%%%%%%%%%%%%%%%%%%%%%%%%

\subsection{Additional Constraints}
\label{ssec:freestreaming}

We have seen in Figure~\ref{couplings} that larger couplings responsible for the decay of $\Phi_{\SM}$ favor lighter dark matter masses. However, light dark matter is constrained because it erases small-scale density perturbations  \cite{Bode:2000gq,Bond:1980ha}. For standard dark matter this bound is $\sim$ 1 keV. Since in FDM models dark matter can be colder, the constraint is relaxed.

When dark matter is relativistic it erases the primordial density perturbations in the matter spectrum, see e.g.~\cite{Bode:2000gq,Bond:1980ha,Gorbunov:2011zz,Lyman}. Therefore the size of the smallest observed gravitationally bound structures probe the horizon size at which dark matter became nonrelativistic. In standard cosmology this horizon size can be related to the temperature and puts a lower bound on the dark matter mass. 
The observation of dwarf galaxies implies density perturbations on comoving scales of $l_{\rm limit}\sim0.1~{\rm Mpc}$ survive. Since density perturbation of order $l_p$ are erased if $l_p \ll l_F$, there is a bound on the free streaming length $l_F$ and, consequently, $m_\DM$. Adapting the treatment in \cite{Gorbunov:2011zz}, we start from the following expression
\beq
l_F=\Big[1+z(T_{\rm NR})\Big]l_H(T_{\rm NR})~,
\eeq
where $[1+z(T_{\rm NR})]\simeq T_{\rm NR}/T_0$ is the redshift at which the dark matter becomes nonrelativistic, in terms of $T_0\approx2.7~{\rm  K} \approx 2.3\times10^{-4}~{\rm eV}$ the present temperature. The horizon size at that time is given by  
\beq
l_H(T_{\rm NR})\equiv \frac{1}{H(T_{\rm NR})} =\left(\frac{90}{g_{\SM}^{\rm (NR)}\pi^2}\right)^{1/2}\frac{M_{\rm Pl}}{T_{\rm NR}^2}~,
\eeq
which implies free streaming length of order
\beq
l_F=\left(\frac{90}{g_{\SM}^{\rm (NR)}\pi^2}\right)^{1/2}
\frac{M_{\rm Pl}}{T_0}\frac{1}{T_{\rm NR}}~.
\label{lf}
\eeq
Now using that FDM becomes nonrelativistic at $T=T_{\rm NR}$ derived in eq.~(\ref{eq:tnr}), we obtain
\beq
l_F=
\left(\frac{90}{g_{\SM}^{\rm (NR)} \pi^2}\right)^{1/2}\frac{M_{\rm Pl}}{m_\DM T_0}
\left(\frac{g_{\SM}^{\rm (NR)}}{g_{\DM}^{(\Gamma)}}\frac{s_\DM^{(\infty)}}{s_\SM^{(\infty)}} \right)^{1/3}~.
\label{eq:FS}
 \eeq
Observe that the free streaming length for FDM is suppressed relative to the expectation for thermal dark matter $l_F^{\rm th}$ as follows  
\beq
\frac{l_F}{l_F^{\rm th}}
\simeq
\left(\frac{m_\DM}{T_{\rm NR}}\right) = \left(\frac{g_{\SM}^{\rm (NR)}}{g_{\DM}^{(\Gamma)}}\frac{s_\DM^{(\infty)}}{s_\SM^{(\infty)}} \right)^{1/3} \simeq 0.12\times\left(\frac{g_{\SM}^{\rm (NR)}}{g_{\DM}^{(\Gamma)}}\frac{1~{\rm keV}}{m_\DM}\right)^{1/3}~.
\label{eq:sup}
 \eeq
Moreover, conservatively requiring that $l_F\lesssim l_{\rm limit}\sim0.1~{\rm Mpc}$ (recall ${\rm Mpc}\approx1.6\times10^{38}~{\rm GeV}^{-1}$), this implies the following lower bound on the dark matter mass, 
\beq
m_\DM\gtrsim 200~{\rm eV},
\label{freestream}
 \eeq
where we take $g_{\DM}^{(\Gamma)}=4$ and $g_{\SM}^{(\rm NR)}\simeq 3.36$ (thus assuming light dark matter for which these bounds are relevant).  This limit is around a factor of 5 weaker than for thermal dark matter.  Note also that we do not weaken the bound arbitrarily because of the additional constraints required for the scenario to be self-consistent. Furthermore, lighter dark matter is not as cold relative to ordinary matter as heavier dark matter would be.

Other experimental constraints on the free streaming length come from Lyman-$\alpha$ \cite{Lyman}  and 21cm line observations \cite{Sitwell:2013fpa,Loeb:2003ya} and these also place bounds on $m_\DM$.  The suppression of the free streaming length relative to the thermal expectation, as given in eq.~(\ref{eq:sup}), implies these constraints are similarly weakened in FDM compared to the bounds on thermal dark matter. Most prominently the current limit from Lyman-$\alpha$ \cite{Lyman} implies a lower limit on thermal dark matter of about 3 keV, thus strengthening the bound by an $\OO(1)$ factor.  For FDM the constraint on the free streaming length can be similarly strengthened by appealing to Lyman-$\alpha$ observations, resulting in the following lower bound (taking $g_{\DM}^{(\Gamma)}=4$)
\beq
m_\DM\gtrsim300~{\rm eV}~.
\label{freestream2}
\eeq
Notice that the dark matter cannot be arbitrarily cold and light  because of the additional constraints for self-consistency of the scenario. In fact the lighter the dark matter, the less the difference in temperature from thermal dark matter.

Moreover, it is anticipated that future 21cm experiments \cite{Tegmark:2008au,Sekiguchi:2014wfa} could improve the lower bound by an order of magnitude and thus probe this scenario up to keV scale dark matter masses. Further complementary probes of light dark matter might the found via analysis of gravitational lensing \cite{Dalal:2002su,Zentner:2003yd,Smith:2011ev} or high-redshift gamma-ray bursts \cite{deSouza:2013wsa}. 

Light dark matter, as permitted by eq.~(\ref{freestream2}), can potentially have observable cosmological consequences if it is relativistic at BBN or last scattering \cite{Cyburt:2015mya,Sarkar:1995dd}. This is typically discussed in terms of additional contributions to $N_{\rm eff}$, the effective number of neutrino species. Given the free-streaming constraints from above we know that dark matter can be relativistic at BBN but not at last scattering, so we only consider constraints from the former. The Standard Model predicts $N_{\rm eff}^{\rm (SM)}=3.046$  \cite{Mangano:2001iu}.  The current $2\sigma$ value inferred from BBN observations (together with data from the CMB and deuterium fraction) is $N_{\rm eff}^{\rm (BBN)}\approx2.9\pm0.4$ \cite{Cyburt:2015mya,Ade:2015xua} and this bounds $\Delta N_{\rm eff}^{\rm (BBN)}\equiv  N_{\rm eff}^{\rm (BBN)} - N_{\rm eff}^{\rm (SM)}<0.25$.
Deviations to $N_{\rm eff}$ due to new relativistic degrees of freedom with energy density $\rho_{\rm rad}$ can be expressed as follows
\beq
\Delta N_{\rm eff}=
\frac{8}{7}\left(\frac{11}{4}\right)^{4/3} \frac{\rho_{\rm rad}}{3\rho_\gamma}~.
\eeq
The ratio of the dark matter and Standard Model energy densities scale together if both are radiation-like. Assuming  dark matter is relativistic at BBN then
$R^{(\Gamma)}\sim R_{\rm BBN}\simeq\frac{\rho_\DM}{\rho_\gamma}\big|_{\rm BBN}$ (this neglects changes to the photon bath from Standard Model states going out of equilibrium). Hence, this leads to a contribution to  $N_{\rm eff}$ proportional to $R^{(\Gamma)}$
\beq
\Delta N_{\rm eff}
&=
\frac{8}{7}\left(\frac{11}{4}\right)^{4/3}\frac{R^{(\Gamma)}}{3}
\sim0.05\left(\frac{4}{g_{\DM}^{(\Gamma)}}\right)^{1/3}\left(\frac{300~{\rm eV}}{m_\DM}\right)^{4/3}~,
\eeq
where the final expressions follows from eq.~(\ref{eq:sos}). This potentially allows for an increase in $N_{\rm eff}$ around the percent level for very light dark matter.

In many models, light degrees of freedom are not a significant concern because the decoupling of heavier Standard Model degrees of freedom reheats Standard Model radiation but not that of a dark sector. However, more generic dark sectors will have heavy decoupling degrees of freedom too, and furthermore there can be several dark sectors. FDM addresses this issue since only the SM degrees of freedom are heated by the entropy dump -- over and above any temperature rise from decoupling. Therefore this dark radiation contributes less to $N_{\rm eff}$.

%%%%%%%%%%%%%%%%%%%%%%%%%%%%%%%%%%%%%%%%%%%

%%%%%%%
\vspace{3mm}
%%%%%%%
\subsection{Allowed Parameter Space}
\label{ssec:parameterspace}

We have outlined a general alternative scenario in which dark matter is present throughout the Universe's evolution, possibly regenerated through decays, and the Standard Model entropy is produced in a late decay.  This scenario is not motivated by any particular coincidence or measurement but by its being part of a very general and probably more likely framework that has not yet been explored. Having looked at the evolution, we have found which parameter space is most plausible and seen that either small couplings or light dark matter (or both) are most promising. Note that successful FDM models must satisfy the following general criteria:
\begin{enumerate}
\item[\bf A.] A thermal bath of $\Phi$ is generated.
\item[\bf B.] The Standard Model reheat temperature is well above BBN.
\item[\bf C.] The relic density of dark matter matches the value observed today.
\end{enumerate}

Condition {\bf A} ensures that a thermal bath of $\Phi$ should be created after inflation, which implies a limit on the mass $m_\Phi\sim \rho_\Phi^{1/4}(a_0)\lesssim 10^{16}$ GeV. This is the anticipated upper bound on the inflaton energy density at reheating in simple models of inflaton \cite{Linde:2005ht}. Furthermore, precision measurements of primordial elements imply that the temperature of the visible sector was in excess of several MeV, before subsequently cooling \cite{Cyburt:2015mya,Sarkar:1995dd}. Condition {\bf B} ensures that these measurements are not perturbed by requiring that $T_{\rm RH}\gtrsim10$ MeV, the temperature of BBN, and this constrains the parameter space through eq.~(\ref{eq:SMrh}) which gives the Standard Model reheat temperature as a function of $\Gamma$. With regards to condition {\bf C}, eq.~(\ref{eq:Ga-0}) gives the form of $\Gamma$ required to match the observed relic density in terms of $m_\Phi$ and $m_\DM$. 

Additionally, as discussed in the introduction the further requirement that baryogenesis occurs, is model dependent. For models in which an asymmetry is generated in leptons and subsequently transferred to baryons via sphalerons \cite{B-via-L}, this requires that the Standard Model is reheated above the electroweak phase transition $T_{\rm EWPT}\sim100$ GeV.

\begin{figure}[t!]
\begin{center}
\includegraphics[height=75mm]{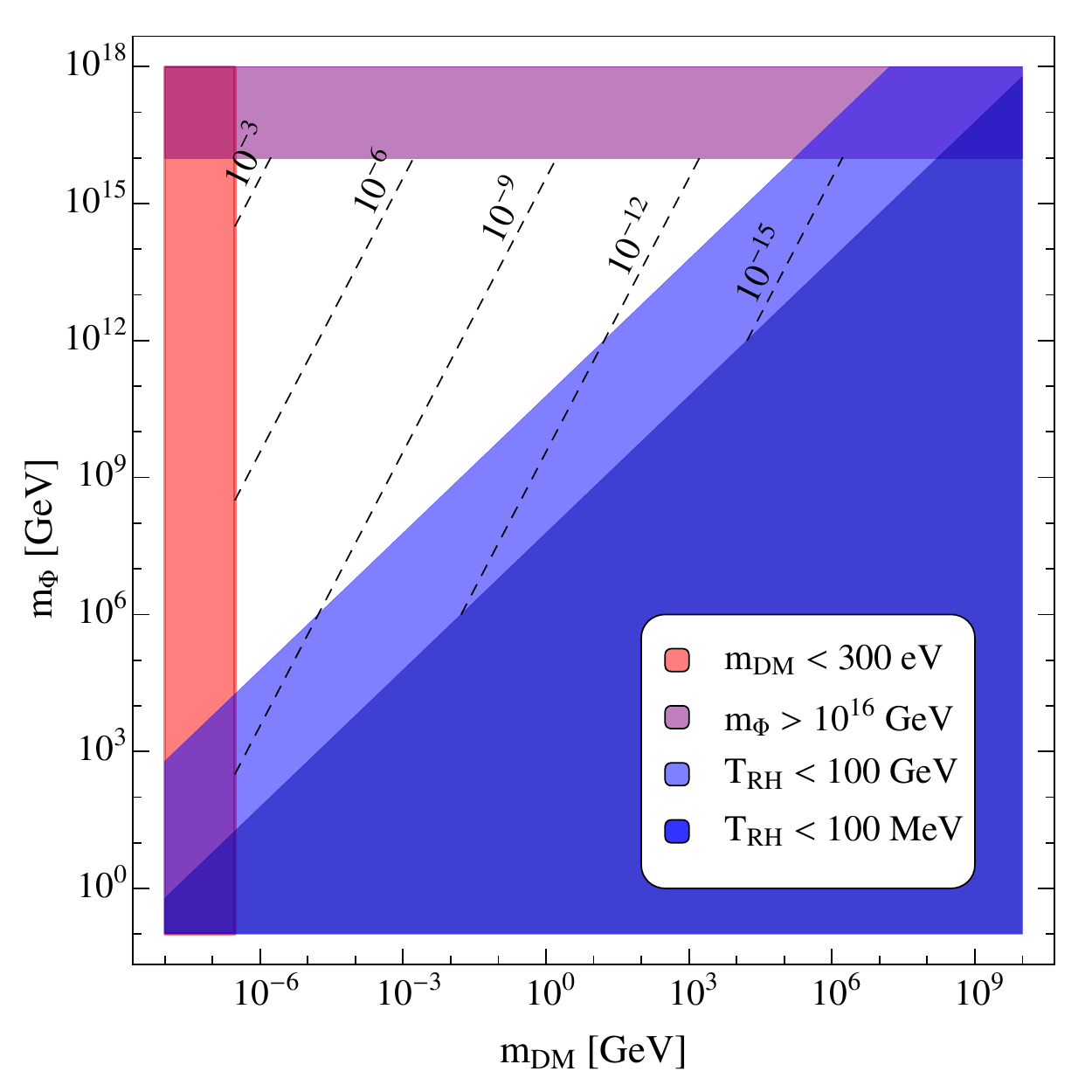}
\caption{
Allowed parameter space for one-field framework in the $m_\DM$--$m_\Phi$ plane, fixed by requiring today's relic energy density of dark matter. The blue regions indicate two reheat temperatures for the visible sector. The red indicates region forbidden by the free streaming constraints from Section~\ref{ssec:freestreaming}. Finally, purple indicates the region in which $\Phi$ would not be populated by the initial inflaton decays. The dashed lines are contours of $\kappa$, defined by $\Gamma = \kappa^2 m_\Phi / 8\pi$.
\label{Fig4}}
\end{center}
\end{figure}

Figure \ref{Fig4} illustrates the available parameter space and the required $\Phi$ decay rate. We present contours of the $\kappa$ necessary to match the observed relic density as a function of $m_\Phi$ and $m_\DM$, for $R^{(0)}=1$. We overlay this with the constraints from BBN ($T_{\rm RH}\gtrsim10$ MeV), free streaming ($m_\DM\gtrsim300$ eV see Section \ref{ssec:freestreaming}), and reheating ($m_\Phi\lesssim10^{16}$ GeV).  In principal a similar constraint plot can be made for models with two heavy states $\Phi_\DM$ and $\Phi_\SM$.

%%%%%%%%%%%%%%%%%%%%%%%%%%%%%%%%%%%%%%%
%%%%%%%%%%%%%%%%%%%%%%%%%%%%%%%%%%%%%%%

%%%%%%%%%%%%%%%%%%%%%%%%%%%%%%%%%%%%%%%%%%
%%%%%%%%%%%%%%%%%%%%%%%%%%%%%%%%%%%%%%%%%%%

\section{Baryogenesis}
\label{sec:baryogenesis}

So far we have focused on the energy and entropy densities stored in ordinary and dark matter. But we have still to discuss baryogenesis and the ratio of baryon number to total Standard Model entropy. The introduction enumerated several distinct mechanisms for generating the baryon asymmetry in FDM models. We examine the first possibility, with a primordial baryon asymmetry, in this section and we will discuss the second option in Section \ref{4.1}.

Scenarios in which the asymmetry is generated prior to $\Phi$ decays are quite attractive since the smallness of the baryon asymmetry might be explained as an $\OO(1)$ asymmetry which is diluted via the late time entropy dump due to $\Phi$ decays. This scenario has the further interesting feature that it puts the baryons and the dark matter on similar footing, because both baryons and dark matter carry similarly low entropy with respect to photons (for dark matter masses above the current bounds). In calculating the late time asymmetry $\Delta^{(\infty)}$ it will be important to account for the relative increase in entropy in the Standard Model sector due to decays of heavy species. For a single decaying heavy state $\Phi$, prior to $\Phi$ decays the entropy of the Standard Model sector and the dark matter sector are related by their ratio of degrees of freedom. After $\Phi$ decays the ratio of entropies in the two sectors stays fixed until today and is given by eq.~(\ref{eq:sos}). We can compare the entropy in each sector before ($s^{(-)}$) and after ($s^{(+)}$) entropy injection. Let us assume that there are no additional entropy dumps into the dark matter sector, thus $s^{(-)}=s^{(+)}$, then it follows that
\beq
\xi \equiv \frac{s_{\SM}^{(-)}}{s_{\SM}^{(+)}} = \frac{s_{\SM}^{(-)}}{s_{\DM}^{(-)}} \frac{s_{\DM}^{(+)}}{s_\SM^{(+)}}  =\frac{g_\SM^{(0)}}{g_\DM^{(0)}}
\frac{2\pi^4}{45\zeta(3)}
\frac{\Delta^{(\infty)}_B \Omega_\DM m_B}{\Omega_B m_\DM}~.
\label{entjump}
\eeq
As a result of entropy injection any initial asymmetry $\Delta^{(0)}$ is diluted to a degree  
\beq
\Delta^{(\infty)}= \xi\Delta^{(0)}~.
\label{eq-5a}
\eeq
Then combining eqs.~(\ref{entjump}) \& (\ref{eq-5a}) and using that for baryons $\Delta^{(\infty)}_B\simeq0.88\times10^{-10}$, we obtain a self-consistency condition
\beq
m_\DM \simeq \frac{2\pi^4}{45\zeta(3)}\frac{g_\SM^{(0)}}{g_\DM^{(0)}} \Delta_B^{(0)}\frac{\Omega_\DM}{\Omega_B} m_B 
\sim
 5~{\rm GeV}~\left(\frac{\Delta_B^{(0)}}{10^{-2}}\right)
\left(\frac{4}{g_\DM^{(\Gamma)}}\right)~.
\label{leptcons0}
\eeq
Conversely, given the dark matter mass, from this expression, the size of the baryon asymmetry prior to dilution can be inferred. We note two interesting extreme cases, according to whether the initial asymmetry takes its maximum or minimum allowed value. 

If the initial asymmetry is maximal, of order $\Delta_B^{(0)}\sim10^{-2}$, a dark matter particle with mass of order a few GeV is favored. This fits in very nicely with asymmetric dark matter \cite{Zurek:2013wia} where the dark matter relic density is also set by a matter-antimatter asymmetry and the baryon and dark matter asymmetries are comparable: $\Delta_B^{(0)}\sim\Delta_\SM^{(0)}$ .  This scenario is much like a conventional asymmetric dark matter model except that the asymmetry is assumed to be produced early on, with a later entropy dump diluting both with respect to photons.

The second case of interest is when the primordial baryon asymmetry is $\Delta_B^{(0)}\sim10^{-9}$ and there is very little entropy injection to the Standard Model. Interestingly, this scenario is compatible with $\sim 500$ eV dark matter particle--at the low end of what is allowed by free-streaming bounds and in a potentially interesting range for solving the core-cusp problem in dwarf galaxies \cite{LJJ}. Of course,  all intermediate mass values are consistent with an appropriate initial asymmetry as determined in eq.~(\ref{leptcons0}). 

It is also of interest to consider the possibility of late $\Phi_{\DM}$ decays which sends entropy into the dark sector, requiring a comparably lighter dark matter for a given initial asymmetry
\beq
m_\DM 
\sim
 5~{\rm MeV}~\left(\frac{\Delta^{(0)}}{10^{-2}}\right)\left(\frac{4}{g_\DM^{(0)}}\right)
 \left(\frac{1000}{s_\DM^{(+)}/s_\DM^{(-)}} \right)
~.
\label{leptcons1}
\eeq
This ratio of dark matter entropies can be expressed parametrically in terms of the model parameters as follows
\beq
\frac{s_\DM^{(+)}}{s_\DM^{(-)}}\sim\left(\frac{\rho^{1/4}_\DM(a_{\Gamma_{\DM}})}{\sqrt{\Gamma_\DM\MP}}\right)^{3}~.
\eeq

%%%%%%%%%%%%%%%%%%%%%%%%%%%%%%%%%%%%%%%%%%
%%%%%%%%%%%%%%%%%%%%%%%%%%%%%%%%%%%%%%%%%%%

\section{See-Saw Neutrino Model}
\label{sec:neutrino}

See-saw neutrino models can generate the correct masses and mixings for neutrinos, as well as potentially account for the generation of lepton number. We will now show that in such models, a heavy right-handed neutrino can play the role of $\Phi$, and furthermore that such models can naturally generate lepton (and hence baryon) number. The left-handed neutrino masses are generated through the operators
\beq
\mathcal{L}_\nu=
y_{ij} H\bar L_iN_j + M_{ij}N_iN_j~.
\label{ref-L}
\eeq
A satisfactory model can be achieved by assuming that all the Yukawa entries have comparable magnitude, which readily explains the large mixing among neutrinos.  
If all entries are comparable, $y^2 v^2/M$ essentially determines the neutrino masses. However, the constraint required for the correct relic abundance depends on the quantity $\Gamma_N (m_\DM/m_N)^2$ as can be seen from eq.~(\ref{eq:Ga-0}), and since $\Gamma_N\simeq y^2 m_N$, this quantity can be expressed $m_\DM^2(y^2/M_N)$. As the factor in brackets is proportional to the neutrino masses, the value of $m_\DM$ is fixed and turns out to be about two orders of magnitude lower than allowed by the Lyman-$\alpha$ bound in our scenario.

However, this analysis assumed that all Yukawa entries are roughly the same (which we assumed only to explain large mixing angles), but this form of the matrix is not essential. The Yukawa matrix can have $\OO(1)$ entries except for one  generation which can have  suppressed couplings:
\begin{equation}
y_{ij}\sim \frac{m_\tau}{v} \times 
\begin{blockarray}{cccc}
N_1 & N_2 & N_3 & \\
\begin{block}{(ccc)c}
  1 & 1 & \epsilon & \nu_1\\
  1 & 1 & \epsilon & \nu_2\\
  1 & 1 & \epsilon & \nu_3\\
\end{block}
\end{blockarray}
\end{equation}
Assuming this coupling structure the left-handed states are generically strongly mixed, as required by observation.

We note that if we take the larger entries of the Yukawa matrix of order the $\tau$ Yukawa coupling, we would have $M\sim10^9~{\rm GeV}$. If we then take the Yukawas associated with the third neutrino much smaller, of order $10^{-6}$ (of order $m_e/m_\tau$), we then see by comparing with Figure \ref{Fig4},  that the rate $\Gamma_{N_3}$ is appropriate to match the dark matter relic density for light dark matter. This can also be see from inspection of eq.~(\ref{eq:sos}); using that $\Gamma_{N_3}\simeq\frac{1}{8\pi} y_e^2 m_{N_3}$, and let us assume $R^{(0)}=1$ and $g_{\rm RH}=100$, one finds
\begin{equation}
\frac{\Omega_\DM}{\Omega_{B}} 
\simeq
\left(\frac{ 45\zeta(3) }{2\pi^4} \right)
\frac{y_{e} m_\DM }{  m_B \Delta}
\sqrt{\frac{\MP}{8\pi M}}
\sim
5\times
\left(\frac{m_\DM}{300~{\rm eV}}\right)
\left(\frac{5\times10^9~{\rm GeV}}{M}\right)^{1/2}~.
\end{equation}
Thus for $y_{\nu_i}\sim y_{l_i}$ and an appropriate choice of $M\sim10^{9}$ GeV one predicts light dark matter of order $m_\DM\sim300$ eV. Some variation in these values is permitted and can be absorbed into $\OO(1)$ changes to $y_{\nu_e}$ relative to $y_{e}$ and the magnitude of $m_{N_3}$. 
Note that this scenario makes a prediction of light neutrino masses, since it makes the lightest neutrino nearly massless and the masses of the other two are solely determined by the measured $\Delta m^2_{ij}$'s. This also predicts that the sum $\Sigma m_\nu $ is as small as can be consistent with the current mass measurements. Although it is possible to test this prediction in the normal hierarchy of the neutrino masses, it would be much harder in the inverted hierarchy, see e.g.~\cite{Abazajian:2013oma}.

%%%%%%%%%%%%%%%%%%%%%%%%%%%%%%%%%%%%%%%%%%%

\subsection{Leptogenesis with Nonthermal Right-handed Neutrino Production}
\label{4.1}

We observe that the Lagrangian of eq.~(\ref{ref-L}) violates $L$ number, and in principle has all the properties required to achieve leptogenesis via $N$ decays \cite{Davidson:2008bu}. Such an asymmetry in the leptons could then  subsequently be transferred to baryons  via sphaleron processes \cite{B-via-L}, provided that the visible sector is reheated above the electroweak phase transition, $T_{\rm RH}\gg 100$ GeV.
We note that there can be more freedom in parameters in this model than in the more conventional leptogenesis models, in which $N$ is produced by thermal production via precisely the same Yukawa couplings that lead to their decay, making for washout of any lepton-generation when processes are in equilibrium. The picture at $H\simeq\Gamma_{N_3}$ bears a resemblance to previously studied models in which leptogenesis proceeds through nonthermal right-handed neutrinos produced via inflation decay \cite{leptog}.
In the usual thermal leptogenesis scenario, the net baryon asymmetry is suppressed by about $10^{-3} \epsilon \eta$ where the first factor is from the large number of Standard Model states that contribute to the net entropy, $\epsilon$ is the net CP violation, and $\eta$ is a factor representing washout and other diluting effects, which in the thermal case is taken to be at most $10^{-1}$. The long lifetime of $N$ implies that at the time of decay right-handed neutrinos are far out of equilibrium and $\eta\approx1$ \cite{Barbieri:1999ma,Buchmuller:2003gz}. Moreover, since $N$ was produced nonthermally the usual constraint on $\eta$ is evaded, and hence the lower bound on $M_N$ is similarly weakened.
  
Of course, violating the usual bound with smaller right-handed neutrino mass would also require smaller Yukawa couplings to generate the known neutrino masses. But in general, asymmetry generation is straightforward. If generated in the final neutrino decay (the decay of the most weakly-coupled neutrino), the asymmetry generated is of order:
\beq
\Delta \sim \left.\frac{n_N}{s_{\SM}}\right|_{\Gamma_{N_3}} \times \epsilon \times \xi \times \eta_{\rm Sph}~,
\label{leptdef}
\eeq
where as already mentioned, $\epsilon$ is the net lepton number generated per decay, $\xi$ is the dilution factor due to entropy generated by the decays of $N$ and $\eta_{\rm Sph} = 8/23$ comes from the sphaleron transfer efficiency~\cite{Harvey:1990qw}.  The quantity $\xi$, the relative increase in entropy in the Standard Model sector due to decays of $N$, is given by eq.~(\ref{entjump}). 
Combining eqs.~(\ref{entjump}) and (\ref{leptdef}) we can derive an analogous condition to eq.~(\ref{leptcons0})
\beq
m_\DM \simeq \frac{2\pi^4}{45\zeta(3)} \frac{g_\SM^{(0)}}{g_\DM^{(0)}} 
 \left.\frac{n_N}{s_{\SM}}\right|_{\Gamma_{N_3}} \epsilon ~ \eta_{\rm Sph}
 \frac{\Omega_\DM}{\Omega_B} m_B ~.
\label{leptcons}
\eeq
Since $\epsilon \lesssim 10^{-4}$, we can only achieve self consistent baryogenesis with $m_\DM\lesssim100$ keV.

%%%%%%%%%%%%%%%%%%%%%%%%%%%%%%%%%%%%%%%%%%%
%%%%%%%%%%%%%%%%%%%%%%%%%%%%%%%%%%%%%%%%%%%

\section{Concluding Remarks}
\label{sec:conclusion}

We have examined the cosmological implications of assuming dark matter arises in an egalitarian fashion along with ordinary matter following inflation, and that there is not necessarily any significant interaction between the different sectors. In particular, we have examined what is necessary to generate the correct entropy and energy ratios for the dark and visible sectors.   Since $\Omega_\DM>\Omega_B$,  without later entropy injection one would expect the dark matter to account for the majority of the entropy.  This problem can be addressed if a heavy state decays into the Standard Model, but leaves the dark matter entropy intact. We do not guarantee that the dark matter energy density is naturally explained, but we do introduce what might be a very generic cosmological scenario with interesting consequences (light dark matter, natural compatibility with baryon asymmetry, and a reasonable implementation in a compelling neutrino scenario) that is worth pursuing.

Given the assumption of democratic inflaton decay,  something must reheat  the Standard Model but not the dark sectors. In FDM the origin of this entropy is the decay of a heavy particle and the present-day ratio of dark matter and baryon abundances is controlled by the lifetimes of such heavy states. FDM models require that the latest decaying heavy state decays into the Standard Model, but not into the dark sector.  This might be likely given the symmetries of the Standard Model if the heavy states are gauge invariant composite operators, in which case we would expect the longest-lived states to be those of lower dimensions, which most likely interact exclusively with only Standard Model or only dark sectors.

In FDM models  the present-day ratio of dark matter and baryon abundances is controlled by the lifetimes of the heavy states $\Phi$. Typically the  final heavy species to decay contributes the most energy and entropy, as earlier energy dumps are diluted relative to the energy in the remaining nonrelativistic states. The last state to decay will typically be the state that is most weakly coupled to its associated sector (i.e.~the state with the longest lifetime). A consequence is that small couplings play a big role in this type of scenario. Indeed, small couplings appear baroque from a model building stance and thus we (playfully) refer to this intriguing selection mechanism as the {\em Maximum Baroqueness Principle}.
This reasoning implies that Standard Model sector is reheated preferentially because it has hierarchically small couplings to the heavy states $\Phi$. Indeed, it is not inconceivable that selection based on maximum baroqueness might be connected with choosing a sector with an electroweak scale hierarchically below the Higgs quadratic cutoff.

One interesting consequence of FDM  is that dark matter can be substantially colder. As a result, its free streaming length is suppressed relative to that expected for thermal dark matter, and thus the lower mass bounds are weakened.  Of particular interest sub-keV fermion dark matter is permitted in FDM potentially offering a minimal resolution of the cusp-core problem due to the importance of Fermi pressure. We will return to this point in a dedicated publication. It is also interesting to note that a light gravitino could be a suitable very light dark matter candidate. A further advantage to this scenario is that it would explain why we don't observe the energy carried by putative light states in dark sectors.

FDM models can also naturally explain why the present-day baryon density is more akin to that of dark matter than to that of the photon bath, as both baryons and dark matter typically contribute similar energy densities and negligible entropy (unless the dark matter is very light). We have shown how FDM can fit in well with either an early baryon asymmetry production or alternatively a later production such as in the neutrino model.  In  the first scenario, according to the dark matter mass, the setting can overlap with that of asymmetric dark matter scenarios.

Since the dark matter is decoupled from the Standard Model, the prospect of observing the state in direct detection or collider experiments is limited.  However, cosmological probes may provide a window: observing, for instance, small deviations to $\Delta N_{\rm eff}$. Additionally, there may be further model dependent probes for a given implementation, such as the prediction of the neutrino mass hierarchy which arises in the heavy right-handed neutrino model of Section \ref{sec:neutrino}. Further, in the fortuitous case that the dark matter is only meta-stable with a lifetime of order the age of the Universe, then one could potentially observe signals of dark matter decays, and a credible signature of decaying dark matter with couplings inconsistent with a thermal relic would provide a strong motivation for the dark matter scenario outlined here.

While we presented an explicit example of this general framework in Section \ref{sec:neutrino} motivated by the neutrino see-saw mechanism, the required small coupling between $\Phi$ and matter particles might arise in a number of alternative scenarios, such as:

%%%%%%%
\vspace{5mm}
%%%%%%%
\begin{itemize}
\item Other models with small technically natural couplings.
\item Kinetic mixing between U(1) mediators \cite{Holdom:1985ag}.
\item  Non-renormalisable operators suppressed by high mass scales (possibly $M_{\rm Pl}$).
\item Non-perturbative effects e.g.~\cite{Kuzmin:1997jua,Carone:2010ha}, with  exponentially suppressed rates, $\Gamma\sim e^{-1/g^2} $, similar to $B+L$ violating decays of $n\rightarrow \bar{p} e^+\bar\nu$ due to electroweak instantons \cite{'tHooft:1976up}.
\end{itemize}
The construction of complete FDM models utilizing small couplings that arise in the manners outlined above would be an interesting continuation of the work initiated here.

In summary, the origin of dark matter abundance can be significantly different than we have probed so far, and can be completely decoupled from Standard Model interactions. Given the genericness of the FDM idea, it is certainly worth deducing the consequences, relating it to existing models, and seeing
whether there are any further possibilities for detection. We have shown this scenario is readily consistent, leading to several interesting consequences. We will pursue the implications for the core-cusp problem and the impact of subsequent freeze-in and freeze-out processes in future publications.

%%%%%%%%%%%%%%%%%%%%%%%%%%%%%%%%%%%%%%%%%%%%
%%%%%%%%%%%%%%%%%%%%%%%%%%%%%%%%%%%%%%%%%%%%

\section*{Acknowledgements} 
We thank W.~Buchm\"uller, D.~Curtin, F.~D' Eramo, L.~Hall, E.~Kramer, and M.~Reece for useful discussions.
This work was supported in part by NSF grants PHY-0855591, PHY-1216270, and PHY-1415548, and the Fundamental Laws Initiative of the Harvard Center for the Fundamental Laws of Nature, and the Munich Institute for Astro- and Particle Physics (MIAPP) of the DFG cluster of excellence ``Origin and Structure of the Universe''.
JS \& JU are also grateful for the hospitality of the CERN theory group.

%%%%%%%%%%%%%%%%%%%%%%%%%%%%%%%%%%%%%%%%%%%%
%%%%%%%%%%%%%%%%%%%%%%%%%%%%%%%%%%%%%%%%%%%%

\appendix

\section{Appendices}
\subsection{Appendix A: Efficiency of Standard Model Reheating}
\label{AppendixA}

In this work we assumed that $\Phi$ suddenly decays into the Standard Model at $3H(a_*) = \Gamma$. However, we know that the decays are gradual and the portion of population of $\Phi$ that decays early has its energy contribution redshifted by the time $3H = \Gamma$. This appendix will investigate the correction that arises from the gradual decay of $\Phi$. In order to quantify this correction we will compare the energy densities of the Standard Model bath in both ``sudden decay'' and ``exact decay'' scenarios at some later time, taken to be $a = 10 a_*$. We will confirm that the sudden decay approximation works well in our setting.

{\bf Sudden Decay:} Assuming $\Phi$ dominates the energy density of the Universe, then $3 M_{\rm Pl}^2 H^2 = m_\Phi n_\Phi$, and in the sudden decay scenario the total energy density at the decay time is $m_\Phi n_\Phi = 3M_{\rm Pl}^2 \Gamma^2_\Phi / \nu^2$. If the initial conditions are $n_\Phi (a=1) = m_\Phi^3$, this happens when
\begin{equation}
a_* = \left(\frac{\nu^2 m_\Phi^4}{3M_{\rm Pl}^2 \Gamma^2} \right)^{1/3}~,
\end{equation}
which fixes the energy density of the Standard Model bath as a function of $a$:
\begin{equation}
\rho_{\rm SM}^{\rm approx}(a) = \frac{3M_{\rm Pl}^2 \Gamma^2_\Phi}{ \nu^2} \left(\frac{a_*}{a}\right)^4 =\nu^{2/3} \left(\frac{m_\Phi^{16}}{3\Gamma^2 M_{\rm pl}^2}\right)^{1/3} a^{-4}~.  
\end{equation}

%%%%%%%%%%%%%%%%%%%%%%%%%%%%%%%%%%%%%%%%%%%
\begin{figure}[t!]
\begin{center}
\includegraphics[height=48mm]{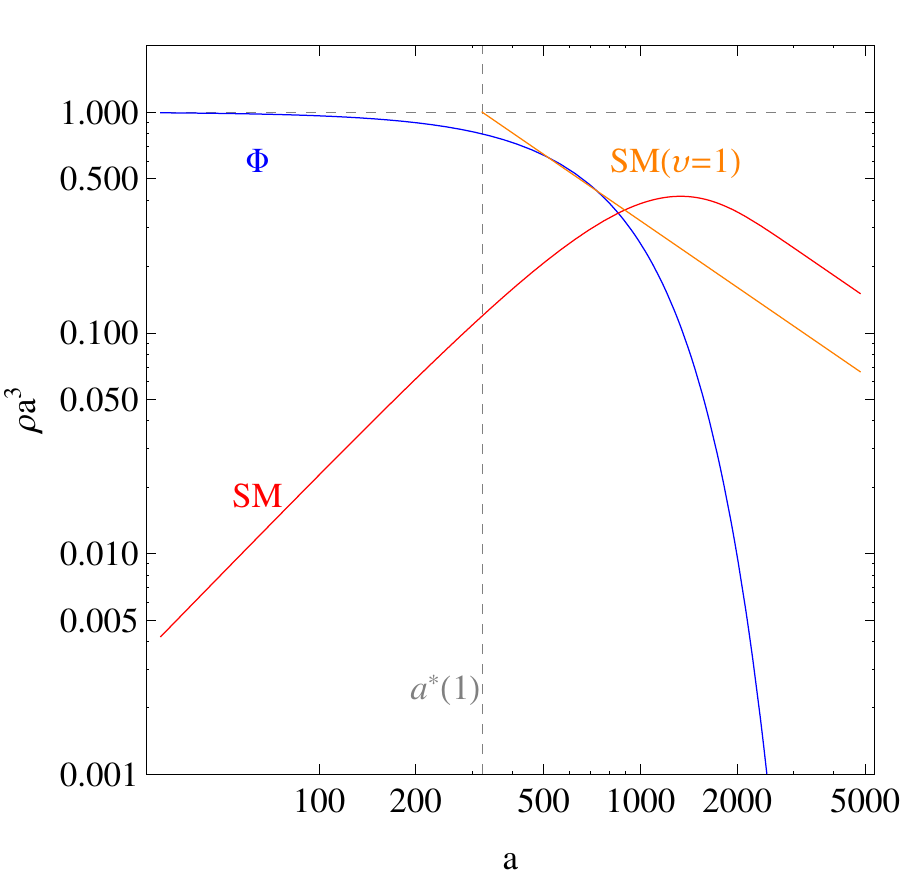}
~
\includegraphics[height=48mm]{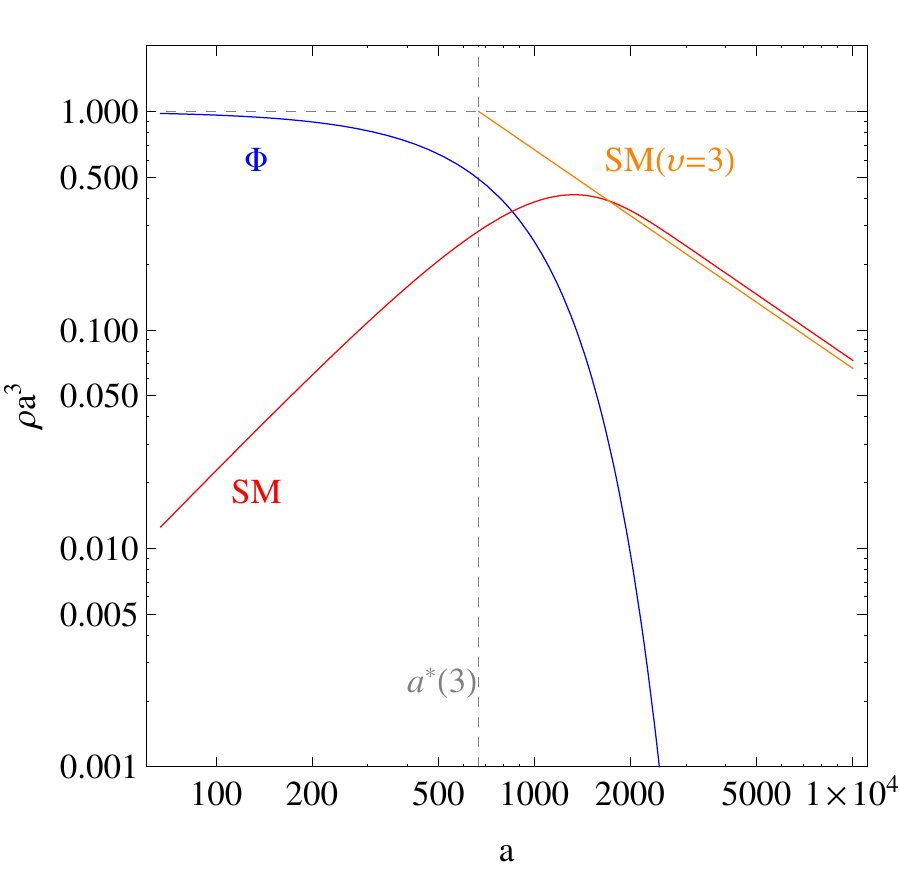}
~
\includegraphics[height=48mm]{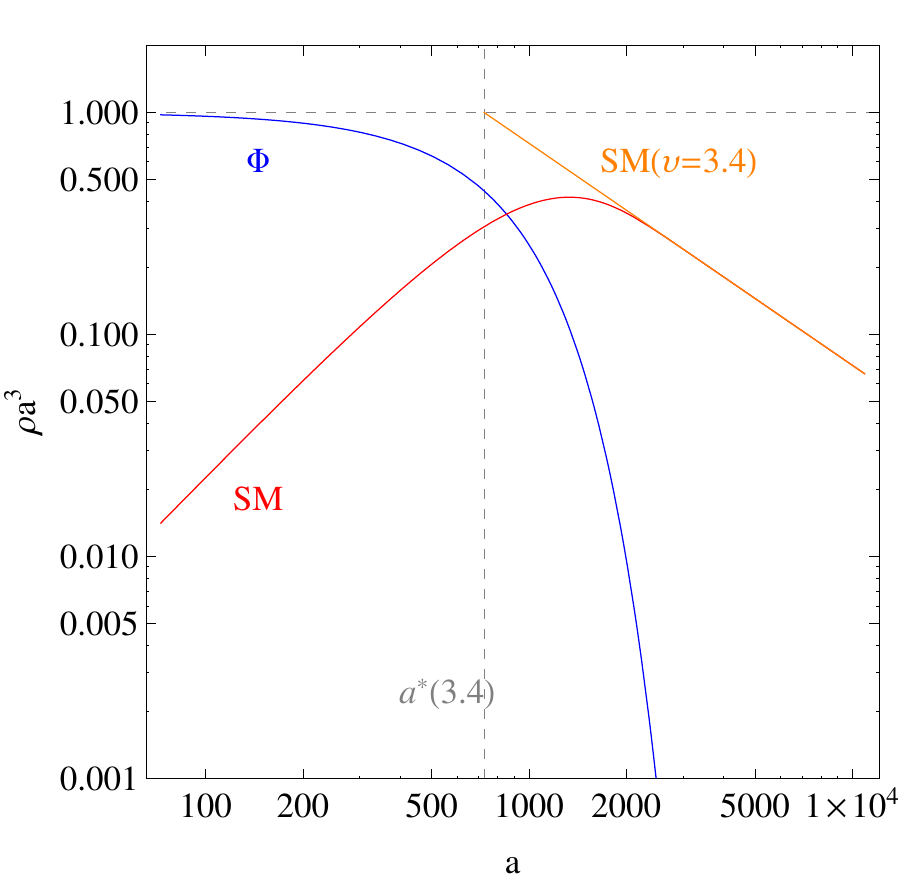}
\caption{Plots show a comparison between the sudden decay approximation and the actual solution for different $\nu$. Observe $\nu=1$  overestimates the reheat energy density by $\sim20\%$, $\nu = 3 $ as used in the text is a good approximation, giving only $\sim2\%$ discrepancy and $\nu=3.4$ gives an excellent match.  
\label{fig:ReheatEfficiency}}
\end{center}
\end{figure}
%%%%%%%%%%%%%%%%%%%%%%%%%%%%%%%%%%%%%%%%%%%

%%%%%%%%%%%%%%%%%%%%%%%%%%%%%%%%%%%%%%%%%%%%

{\bf Exact Decay:} To assess the process of energy transfer from $\Phi$ to the Standard Model in the exact solution, we set up a system of differential equations that track the evolution of the number density $n$ of $\Phi$'s and energy density of the Standard Model bath $\rho$
\beq
\dot{n} + 3H n  &= -  \Gamma n~,
\hspace{15mm}
\dot{\rho} + 4H \rho &= \Gamma m_\Phi n~,
\eeq
where dotted variables indicate a derivative w.r.t.~regular time and $H^2 = (m_\Phi n_\Phi+\rho)/3M_{\rm Pl}^2$. It is convenient to rewrite the above in terms of derivatives with respect to $a$, which we denote by primed variables. Note that in general $\dot{x} = a H x'$ and it follows that
\beq
a H n' + 3H n  &= -  \Gamma n~,
\hspace{15mm}
a H \rho' + 4H \rho &= \Gamma m_\Phi n~.
\eeq
We take the same initial conditions as for the sudden decay scenario $n_0(a=1)= m_\Phi^3$ and $\rho=0$, and we choose $m_\Phi=10^{-2} M_{\rm pl}$ and $\Gamma = 10^{-6} m_\Phi$, then we plot the energy density in the Standard Model sector as a function of $a$ in Figure \ref{fig:ReheatEfficiency}.

%%%%%%%%%%%%%%%%%%%%%%%%%%%%%%%%%%%%%%%%%%%%

{\bf Comparison:}
Taking the same values for $m_\Phi$ and $\Gamma$ for the sudden decay as used in the exact decay case allows us to compare the efficiency of energy transfer to the Standard Model sector. For $\nu=1$, we get $\Theta \equiv \rho_{\rm approx}/\rho_{\rm exact} = 1.22$, with $\nu = 3$, we get $\Theta = 1.021$, finally, for $\nu = 3.4$ one has $\Theta = 1.00$. The quality of the approximation can be seen from inspection of Figure \ref{fig:ReheatEfficiency}.  Moreover, these results are robust and hold very close over a range of $m_\Phi$ and $\Gamma$. Whilst $\nu \approx 3.4$ is the best match, $\nu = 3$ used in the main text provides a very good approximation.

%%%%%%%%%%%%%%%%%%%%%%%%%%%%%%%%%%%%%%%%%%%%
%%%%%%%%%%%%%%%%%%%%%%%%%%%%%%%%%%%%%%%%%%%%

\subsection{Appendix B: Nonrelativistic Dark Matter Prior to Reheating}
\label{AppendixB}

In this appendix we examine the case in which the dark matter becomes relativistic before $\Phi$ decays and show that the decay rate required to match the observed relic is of a highly similar form to the converse scenario studied in Section \ref{ssec:onefield}. At the point of decay, by definition, the dark matter mass density is given by
\beq
m_\DM n_\DM(a_{\Gamma}) = \rho_\DM(a_{\Gamma}) = R_{\rm \Gamma}  \rho_\Phi(a_{\Gamma})~.
\eeq
It will be useful to introduce the quantity $m_B n_{B-\bar{B}}$, being the projected baryon mass density duet the late time asymmetry. By evaluating this quantity after $\Phi$ decays as follows
\beq
m_B n_{B-\bar{B}}(a_{\Gamma}) = m_B \Delta s_\SM 
= m_B \Delta \frac{4}{3T_{\rm RH}}\rho_{\Phi}(a_{\Gamma})~.
\eeq
we can readily obtain a relation for the asymptotic ratio of mass densities at the present-day
\beq
\frac{\Omega_\DM}{\Omega_{B}} =
\frac{m_\DM n_\DM(a_{\Gamma}) }{m_B n_{B-\bar{B}}(a_{\Gamma})}
= \frac{3R_{\rm \Gamma}}{4m_B \Delta} ~\sqrt{\Gamma\MP}
\left(\frac{10}{\pi^2 g_{\rm SM}^{(\Gamma)} }\right)^{1/4}~,
\label{eq:omega}
\eeq
where we use the expression for $T_{\rm RH}$ from eq.~(\ref{eq:SMrh}). 
 We now require an expression for $R_\Gamma$ in the case that the dark matter is nonrelativistic prior to $\Phi$ decays. The $X$ and $\Phi$ densities redshift relative to each other only until both species become nonrelativistic and thus $R_{\rm NR}=R_\Gamma$.

At $a=a_0$ the energy density of the dark matter sector is  $\rho_\DM(a_0)\simeq m_{\Phi}^4$  whereas at the point that the dark matter becomes nonrelativistic  $\rho_\DM(a_{\rm NR})\simeq m_\DM^4$. As a result:
\beq
\frac{a_0}{a_{\rm NR}}=\left(\frac{\rho_\DM(a_{\rm NR})}{\rho_\DM(a_{0})}\right)^{1/4}= \frac{m_\DM}{m_{\Phi}} \left(\frac{1}{R_0} \right)^{1/4}~,
\eeq
Since $\Phi$ is nonrelativistic its energy density of $\Phi$ at the point the dark matter becomes nonrelativistic is just
\beq
\rho_{\Phi}(a_{\rm RN})= \rho_{\Phi}(a_{0})  \left(\frac{a_{0}}{a_{\rm RN}} \right)^3=
\frac{g_\Phi \pi^2}{30} m_\DM^3m_\Phi  \left(\frac{1}{R_0} \right)^{3/4} ~.
\label{eq:phi}
\eeq
Once the dark matter is nonrelativistic $\Phi$ and $X$ redshift at the same rate until $\Phi$ decays, hence
\beq
R_\Gamma=R_{\rm NR}\equiv\frac{\rho_\DM(a_{\rm NR})}{\rho_\Phi(a_{\rm NR})} 
=\frac{g_\DM}{g_\Phi}\frac{m_\DM}{m_\Phi} \left(R_0\right)^{3/4}~.
\label{eq:rnr}
\eeq
It follows from eq.~(\ref{eq:omega}) and (\ref{eq:rnr}) that the required decay rate is
\beq
 \Gamma 
 &= 
 \frac{\pi}{\sqrt{10}}
 \frac{m_\Phi^2}{M_{\rm Pl}}
 \left[\frac{4}{3}\Delta \frac{ \Omega_\DM}{\Omega_B}\frac{m_B}{m_\DM}\right]^2 \left(\frac{g_\Phi}{g_\DM}\right)^2  \sqrt{\frac{g_\SM^{(\Gamma)}}{\left(R^{(0)}\right)^3}}~.
 \eeq
Note that the factor in square brackets is the same as the ratio of entropies given in eq.~(\ref{eq:sos}). Thus up to numerical prefactors this result is the same as eq.~(\ref{eq:Ga-0}), which was derived under the converse assumption that the dark matter remained relativistic until after $H\sim\Gamma$.
%%%%%%%%%%%%%%%%%%%%%%%%%%%%%%%%%%%%%%%%%%%%

%%%%%%%%%%%%%%%%%%%%%%%%%%%%%%%%%%%%%%%%%%
%%%%%%%%%%%%%%%%%%%%%%%%%%%%%%%%%%%%%%%%%%%

\subsection{Appendix C: Multiple Non-degenerate Heavy States}
\label{ApC}

Let us assume the ordering of parameters $m_{\Phi_\SM}>m_{\Phi_\DM}$ and $\Gamma_{\DM}>\Gamma_{\SM}$. The mass threshold of $\Phi_{\SM}$ is the earliest distinguished cosmological marker and thus we now define $a_0\equiv a(T=m_{\Phi_\DM})$. At this point all states are relativistic and we take the initial conditions 
$\rho_i(a_0)=R^{(i)}_0 m_{\Phi_\SM}^4~,$ the $R^{(i)}_0$ account for the initial ratios of $i=\Phi_\DM,\Phi_\SM,$ DM, SM.

The energy densities are evolved to the $\Phi_{\DM}$ mass threshold marked by $a_{\Phi_\DM}$
\beq
\rho_{\Phi_\DM}(a_{\Phi_\DM}) &=g_{\Phi_\SM}R^{(\Phi_\DM)}_0 m_{\Phi_\SM}^4\left(\frac{a_0}{a_{\Phi_\DM}}\right)^4,\\
\rho_{\Phi_\SM}(a_{\Phi_\DM}) &=g_{\Phi_\DM}R^{(\Phi_\SM)}_0 m_{\Phi_\SM}^4\left(\frac{a_0}{a_{\Phi_\DM}}\right)^3.
\eeq
We neglect to track the dark matter and Standard Model as these will be subsequently replenished via decays of $\Phi_i$. The redshift factor is determined by the evolution of Hubble, as in Section \ref{ssec:onefield}.
Evolution from the $\Phi_{\SM}$ mass threshold to the $\Phi_{\DM}$ mass threshold assuming matter domination is governed by
\beq
\left(\frac{a_0}{a_{\Phi_\DM}}\right)^3
= \frac{1}{R^{(\Phi_\SM)}_0}\frac{g_{\Phi_\DM}}{g_{\Phi_\SM}}\left(\frac{m_{\Phi_\DM}}{m_{\Phi_\SM}}\right)^4~.
\eeq
Thus the energy densities evolve to
\beq
\rho_{\Phi_\SM}(a_{\Phi_\DM}) &
=\left(\frac{g_{\Phi_\DM}}{g_{\Phi_\SM}}\right)^{4/3}\left[\frac{R^{(\Phi_\DM)}_0}{R^{(\Phi_\SM)}_0}\left(\frac{1}{R^{(\Phi_\SM)}_0}\frac{m_{\Phi_\DM}}{m_{\Phi_\SM}}\right)^{1/3}\right]m_{\Phi_\DM}^4
\equiv r \left(\frac{g_{\Phi_\DM}}{g_{\Phi_\SM}}\right)m_{\Phi_\DM}^4\\
\rho_{\Phi_\DM}(a_{\Phi_\DM}) &= \left(\frac{g_{\Phi_\DM}}{g_{\Phi_\SM}}\right)m_{\Phi_\DM}^4~.
\eeq
Now we evolve to $H=\Gamma_{\DM}/3$ marked by $a_{\Gamma_{\DM}}$. Assuming that at  $H\simeq\Gamma_{\DM}$ the Universe is matter dominated, the redshift factor is 
\beq
\left(\frac{a_{\Phi_\DM}}{a_{\Gamma_{\DM}}}\right)^3=
\frac{3}{1+r}
\left(\frac{g_{\Phi_\SM}}{g_{\Phi_\DM}}\right)
\left(\frac{\Gamma_{\DM}^2M_{\rm Pl}^2}{m_{\Phi_\DM}^4}\right)~.
\eeq
As both $\Phi$ states are redshifting as matter during this stage, the energy densities are
\beq
\rho_{\Phi_\SM}(a_{\Gamma_{\DM}}) &=
  3\Gamma_{\DM}^2M_{\rm Pl}^2
  \left(\frac{1}{1+r}\right) ~,
\hspace{10mm}
\rho_{\Phi_\DM}(a_{\Gamma_{\DM}}) &= r\rho_{\Phi_\SM}(a_{\Gamma_{\DM}}) ~.
\eeq
Moreover we make the identification $\rho_\DM(a_{\Gamma_{\DM}})=\rho_{\Phi_{\DM}}(a_{\Gamma_{\DM}})$.
Finally we evolve to $H=\Gamma_{\SM}/3$, assuming $\Phi_{\SM}$ decays once matter domination is restored following the decay of $\Phi_{\DM}$
\beq
 \left(\frac{a_{\Gamma_{\DM}}}{a_{\Gamma_{\SM}}}\right)^3=
 (1+r)\left(\frac{\Gamma_{\SM}}{\Gamma_{\DM}}\right)^2 
~.
\eeq
Then the dark matter and Standard Model energy densities at $3H=\Gamma_{\SM}$ decay are given by
\beq
\rho_\SM(a_{\Gamma_{\SM}}) 
&=\rho_{\Phi_\DM}(a_{\Gamma_{\DM}})\left(\frac{a_{\Gamma_{\DM}}}{a_{\Gamma_{\SM}}}\right)^3  = 3\Gamma_{\SM}^2M_{\rm Pl}^2 \\
\rho_\DM(a_{\Gamma_{\SM}}) &=\rho_{\Phi_\SM}(a_{\Gamma_{\DM}}) \left(\frac{a_{\Gamma_{\DM}}}{a_{\Gamma_{\SM}}}\right)^4=
3\Gamma_{\SM}^2M_{\rm Pl}^2
~r\left( (1+r)\left(\frac{\Gamma_{\SM}}{\Gamma_{\DM}}\right)^2 \right)^{1/3}~.
\eeq
From the equations above one can derive an analogous expressions eq.~(\ref{eq-n1}) and subsequently the required $\Gamma_{\SM}$ for a given set of parameters ($\Gamma_{\DM}$, $m_\DM$, $m_{\Phi_\SM}$, $m_{\Phi_\DM}$) in order to match the observed relic density, similar to eq.~(\ref{fi-1}).


\vspace{-2mm}\begin{thebibliography}{}   


\bibitem{moduli}
%\bibitem{Banks:1993en}
  T.~Banks, D.~B.~Kaplan and A.~E.~Nelson,
{\em Cosmological implications of dynamical supersymmetry breaking,}
  Phys.\ Rev.\ D {\bf 49} (1994) 779
  [hep-ph/9308292].
  %%CITATION = HEP-PH/9308292;%%
  %
%\bibitem{de Carlos:1993jw}
  B.~de Carlos, J.~A.~Casas, F.~Quevedo and E.~Roulet,
  {\em Model independent properties and cosmological implications of the dilaton and moduli sectors of 4-d strings,}
  Phys.\ Lett.\ B {\bf 318} (1993) 447
  [hep-ph/9308325].
  %%CITATION = HEP-PH/9308325;%%
   %
  %\bibitem{Gelmini:2006pw}
  G.~B.~Gelmini and P.~Gondolo,
  {\em Neutralino with the right cold dark matter abundance in (almost) any supersymmetric model,}
  Phys.\ Rev.\ D {\bf 74} (2006) 023510
  [hep-ph/0602230].
    %%CITATION = doi:10.1103/PhysRevD.74.023510;%%
%
%\bibitem{Hui:1998dc}
  L.~Hui and E.~D.~Stewart,
  {\em Superheavy dark matter from thermal inflation,}
  Phys.\ Rev.\ D {\bf 60} (1999) 023518
  [hep-ph/9812345].
  %%CITATION = doi:10.1103/PhysRevD.60.023518;%%

\bibitem{Moroi:1999zb}
  T.~Moroi and L.~Randall,
  {\em Wino cold dark matter from anomaly mediated SUSY breaking,}
  Nucl.\ Phys.\ B {\bf 570} (2000) 455
  [hep-ph/9906527].
  %%CITATION = HEP-PH/9906527;%%


        \bibitem{LJJ}
L.~Randall, J.~Scholtz and J.~Unwin, in preparation.
    
\bibitem{inflaton-baryo}
%\bibitem{Affleck:1984fy}
  I.~Affleck and M.~Dine,
  {\em A New Mechanism for Baryogenesis,}
  Nucl.\ Phys.\ B {\bf 249} (1985) 361.
  %%CITATION = NUPHA,B249,361;%%
  %
  %\cite{Linde:1985gh}
%\bibitem{Linde:1985gh}
  A.~D.~Linde,
  {\em The New Mechanism of Baryogenesis and the Inflationary Universe,}
  Phys.\ Lett.\ B {\bf 160} (1985) 243.
  %%CITATION = PHLTA,B160,243;%%
  %
%\bibitem{Delepine:2006rn}
  D.~Delepine, C.~Martinez and L.~A.~Urena-Lopez,
   {\em Complex Hybrid Inflation and Baryogenesis,}
  Phys.\ Rev.\ Lett.\  {\bf 98} (2007) 161302
  [hep-ph/0609086].
  %%CITATION = HEP-PH/0609086;%%
 %
  %\bibitem{BasteroGil:2011cx}
  M.~Bastero-Gil, A.~Berera, R.~O.~Ramos and J.~G.~Rosa,
  {\em Warm baryogenesis,}
  Phys.\ Lett.\ B {\bf 712} (2012) 425
  [1110.3971].
  %%CITATION = ARXIV:1110.3971;%%
  %
 %\bibitem{Hertzberg:2013mba}
  M.~P.~Hertzberg and J.~Karouby,
   {\em Generating the Observed Baryon Asymmetry from the Inflaton Field,}
  Phys.\ Rev.\ D {\bf 89} (2014) 6,  063523
  [1309.0010].
  %%CITATION = ARXIV:1309.0010;%%
  %
%\bibitem{Hook:2014mla}
  A.~Hook,
   {\em Baryogenesis from Hawking Radiation,}
  Phys.\ Rev.\ D {\bf 90} (2014) 8,  083535
  [1404.0113].
  %%CITATION = ARXIV:1404.0113;%%
  %
%\bibitem{Barrie:2014waa}
  N.~D.~Barrie and A.~Kobakhidze,
   {\em Inflationary Baryogenesis in a Model with Gauged Baryon Number,}
  JHEP {\bf 1409} (2014) 163
  [1401.1256].
  %%CITATION = ARXIV:1401.1256;%%
 %
% \bibitem{Unwin:2015wya}
 J.~Unwin,
  {\em On Baryogenesis from a Complex Inflaton,}
  [1503.06806].
  %%CITATION = ARXIV:1503.06806;%%
  %
%\bibitem{Hook:2015foa}
  A.~Hook,
  {\em Baryogenesis in a CP invariant theory,}
  [1508.05094].
    %%CITATION = ARXIV:1508.05094;%%

%%%%%%

\bibitem{Davidson:2008bu}
    See e.g.~S.~Davidson, E.~Nardi and Y.~Nir,
{\em Leptogenesis,}
  Phys.\ Rept.\  {\bf 466} (2008) 105
  [0802.2962].
  %%CITATION = ARXIV:0802.2962;%%
  

    \bibitem{leptog}
%\bibitem{Lazarides:1991wu}
  G.~Lazarides and Q.~Shafi,
{\em Origin of matter in the inflationary cosmology,}
  Phys.\ Lett.\ B {\bf 258} (1991) 305.
  %%CITATION = PHLTA,B258,305;%%
  %
%  \bibitem{Giudice:1999fb}
  G.~F.~Giudice, M.~Peloso, A.~Riotto and I.~Tkachev,
{\em Production of massive fermions at preheating and leptogenesis,}
  JHEP {\bf 9908} (1999) 014
  [hep-ph/9905242].
  %%CITATION = HEP-PH/9905242;%%
  %
%\bibitem{Buchmuller:2005eh}
W.~Buchmuller, R.~D.~Peccei and T.~Yanagida,
 {\em Leptogenesis as the origin of matter,}
  Ann.\ Rev.\ Nucl.\ Part.\ Sci.\  {\bf 55} (2005) 311
  [hep-ph/0502169].
  %%CITATION = HEP-PH/0502169;%%

 \bibitem{other}
 %\bibitem{Dine:1995kz}
  M.~Dine, L.~Randall and S.~D.~Thomas,
  {\em Baryogenesis from flat directions of the supersymmetric standard model,}
  Nucl.\ Phys.\ B {\bf 458} (1996) 291
  [hep-ph/9507453].
  %%CITATION = doi:10.1016/0550-3213(95)00538-2;%%
%
%  \bibitem{Thomas:1995ze} 
  S.~D.~Thomas,
  {\em Baryons and dark matter from the late decay of a supersymmetric condensate,}
  Phys.\ Lett.\ B {\bf 356}, 256 (1995)
  [hep-ph/9506274].
  %%CITATION = doi:10.1016/0370-2693(95)00772-D;%%
 %
  R.~Allahverdi, B.~Dutta and K.~Sinha,
  {\em Baryogenesis and Late-Decaying Moduli,}
  Phys.\ Rev.\ D {\bf 82} (2010) 035004
  [1005.2804].
  %%CITATION = doi:10.1103/PhysRevD.82.035004;%%
%\\
%\bibitem{Unwin:2012rp}
  J.~Unwin,
  {\em Exodus: Hidden origin of dark matter and baryons,}
  JHEP {\bf 1306} (2013) 090
  [1212.1425].
  %%CITATION = doi:10.1007/JHEP06(2013)090;%%
%  \\
%  \bibitem{Perez:2013nra}
  P.~Fileviez Perez and M.~B.~Wise,
 {\em Baryon Asymmetry and Dark Matter Through the Vector-Like Portal,}
  JHEP {\bf 1305} (2013) 094
  [1303.1452].
  %%CITATION = doi:10.1007/JHEP05(2013)094;%%
%\\ 
%\bibitem{Kane:2015jia}
  G.~Kane, K.~Sinha and S.~Watson,
  {\em Cosmological Moduli and the Post-Inflationary Universe: A Critical Review,}
  Int.\ J.\ Mod.\ Phys.\ D {\bf 24} (2015) 08,  1530022
  [1502.07746].
  %%CITATION = doi:10.1142/S0218271815300220;%%


%%%%%%

\bibitem{Morrissey:2012db}
For a review see e.g.~D.~E.~Morrissey and M.~J.~Ramsey-Musolf,
  {\em Electroweak baryogenesis,}
  New J.\ Phys.\  {\bf 14} (2012) 125003
  [1206.2942].
  %%CITATION = ARXIV:1206.2942;%%
  
\bibitem{FI}
%\bibitem{Hall:2009bx}
  L.~J.~Hall, K.~Jedamzik, J.~March-Russell and S.~M.~West,
  {\em Freeze-In Production of FIMP Dark Matter,}
  JHEP {\bf 1003} (2010) 080
  [0911.1120].
  %%CITATION = ARXIV:0911.1120;%%
%
 % \bibitem{McDonald:2001vt}
  J.~McDonald,
  {\em Thermally generated gauge singlet scalars as self-interacting dark matter,}
  Phys.\ Rev.\ Lett.\  {\bf 88} (2002) 091304
  [hep-ph/0106249];
  %%CITATION = HEP-PH/0106249;%%
  %
 %\bibitem{Cheung:2010}
  C.~Cheung, G.~Elor, L.~J.~Hall and P.~Kumar,
  {\em Origins of Hidden Sector Dark Matter I: Cosmology,}
  JHEP {\bf 1103} (2011) 042
  [1010.0022].
  %%CITATION = ARXIV:1010.0022;%%
  %
%\bibitem{Chu:2011be}
  X.~Chu, T.~Hambye and M.~Tytgat,
   {\em The Four Basic Ways of Creating Dark Matter Through a Portal,}
  JCAP {\bf 1205} (2012) 034
  [1112.0493];
  %%CITATION = ARXIV:1112.0493;%%
%  
  % \bibitem{Blennow:2013jba}
  M.~Blennow, E.~Fernandez-Martinez and B.~Zaldivar,
  {\em Freeze-in through portals,}
  JCAP {\bf 1401} (2014) 01,  003
  [1309.7348];
  %%CITATION = ARXIV:1309.7348;%%
  %
%  \bibitem{Dev:2013yza}
  P.~S.~Bhupal Dev, A.~Mazumdar and S.~Qutub,
  {\em Constraining Non-thermal and Thermal properties of Dark Matter,}
  Physics {\bf 2} (2014) 26
  [1311.5297];
  %%CITATION = ARXIV:1311.5297;%%
  %
%\bibitem{Elahi:2014fsa}
  F.~Elahi, C.~Kolda and J.~Unwin,
  {\em UltraViolet Freeze-in,}
  JHEP {\bf 1503} (2015) 048
  [1410.6157].
  %%CITATION = ARXIV:1410.6157;%%
   %
%\bibitem{Roland:2014vba}
  S.~B.~Roland, B.~Shakya and J.~D.~Wells,
   {\em Neutrino Masses and Sterile Neutrino Dark Matter from the PeV Scale,}
  [1412.4791].
  %%CITATION = ARXIV:1412.4791;%%
   %
%  \bibitem{Co:2015pka}
  R.~T.~Co, F.~D'Eramo, L.~J.~Hall and D.~Pappadopulo,
  {\em Freeze-In Dark Matter with Displaced Signatures at Colliders,}
 [1506.07532].
  %%CITATION = ARXIV:1506.07532;%%
  
\bibitem{Sarkar:1995dd}
  For a review see e.g.~S.~Sarkar,
   {\em Big bang nucleosynthesis and physics beyond the standard model,}
  Rept.\ Prog.\ Phys.\  {\bf 59} (1996) 1493
  [hep-ph/9602260].
  %%CITATION = HEP-PH/9602260;%%
  
\bibitem{Cyburt:2015mya}
  R.~H.~Cyburt, B.~D.~Fields, K.~A.~Olive and T.~H.~Yeh,
  {\em Big Bang Nucleosynthesis: 2015,}
  [1505.01076].
  %%CITATION = ARXIV:1505.01076;%%
  
\bibitem{Bond:1980ha}
  J.~R.~Bond, G.~Efstathiou and J.~Silk,
  {\em Massive Neutrinos and the Large Scale Structure of the Universe,}
  Phys.\ Rev.\ Lett.\  {\bf 45} (1980) 1980.
  %%CITATION = PRLTA,45,1980;%%
  
\bibitem{Bode:2000gq}
  P.~Bode, J.~P.~Ostriker and N.~Turok,
 {\em Halo formation in warm dark matter models,}
  Astrophys.\ J.\  {\bf 556} (2001) 93
  [astro-ph/0010389].
  %%CITATION = ASTRO-PH/0010389;%%
    
            
\bibitem{Gorbunov:2011zz}
  D.~S.~Gorbunov and V.~A.~Rubakov,
{\em Introduction to the theory of the early universe: Hot big bang theory,}
World Scientific (2011).
  %%CITATION = INSPIRE-963505;%%


\bibitem{Lyman}
%\bibitem{Narayanan:2000tp}
  V.~K.~Narayanan, D.~N.~Spergel, R.~Dave and C.~P.~Ma,
{\em Constraints on the mass of warm dark matter particles and the shape of the linear power spectrum from the Ly$\alpha$ forest,}
  Astrophys.\ J.\  {\bf 543} (2000) L103
  [astro-ph/0005095];
  %%CITATION = ASTRO-PH/0005095;%%
%
 % \bibitem{Bolton:2004ge}
  J.~S.~Bolton, M.~G.~Haehnelt, M.~Viel and V.~Springel,
  {\em The Lyman-alpha forest opacity and the metagalactic hydrogen ionization rate at z $\sim$ 2-4,}
  Mon.\ Not.\ Roy.\ Astron.\ Soc.\  {\bf 357} (2005) 1178
  [astro-ph/0411072];
  %%CITATION = ASTRO-PH/0411072;%%
%
%\bibitem{Viel:2013fqw}
  M.~Viel, G.~D.~Becker, J.~S.~Bolton and M.~G.~Haehnelt,
  {\em Warm dark matter as a solution to the small scale crisis: New constraints from high redshift Lyman-$\alpha$ forest data,}
  Phys.\ Rev.\ D {\bf 88} (2013) 043502
  [1306.2314].
  %%CITATION = ARXIV:1306.2314;%%

\bibitem{Loeb:2003ya}
  A.~Loeb and M.~Zaldarriaga,
  {\em Measuring the small - scale power spectrum of cosmic density fluctuations through 21 cm tomography prior to the epoch of structure formation,}
  Phys.\ Rev.\ Lett.\  {\bf 92} (2004) 211301
  [astro-ph/0312134].
  %%CITATION = ASTRO-PH/0312134;%%
  
\bibitem{Sitwell:2013fpa}
  M.~Sitwell, A.~Mesinger, Y.~Z.~Ma and K.~Sigurdson,
  {\em The Imprint of Warm Dark Matter on the Cosmological 21-cm Signal,}
  Mon.\ Not.\ Roy.\ Astron.\ Soc.\  {\bf 438} (2014) 3,  2664
  [1310.0029].
  %%CITATION = ARXIV:1310.0029;%%

\bibitem{Sekiguchi:2014wfa}
  T.~Sekiguchi and H.~Tashiro,
{\em Constraining warm dark matter with 21 cm line fluctuations due to minihalos,}
  JCAP {\bf 1408} (2014) 007
  [1401.5563].
  %%CITATION = ARXIV:1401.5563;%%
  
  \bibitem{Tegmark:2008au}
  M.~Tegmark and M.~Zaldarriaga,
 {\em The Fast Fourier Transform Telescope,}
  Phys.\ Rev.\ D {\bf 79} (2009) 083530
  [0805.4414].
  %%CITATION = ARXIV:0805.4414;%%
  
\bibitem{Dalal:2002su}
  N.~Dalal and C.~S.~Kochanek,
  {\em Strong lensing constraints on small scale linear power,}
  [astro-ph/0202290].
  %%CITATION = ASTRO-PH/0202290;%%

 \bibitem{Zentner:2003yd}
  A.~R.~Zentner and J.~S.~Bullock,
  {\em Halo substructure and the power spectrum,}
  Astrophys.\ J.\  {\bf 598} (2003) 49
  [astro-ph/0304292].
  %%CITATION = ASTRO-PH/0304292;%%
 
 \bibitem{Smith:2011ev}
  R.~E.~Smith and K.~Markovic,
 {\em Testing the Warm Dark Matter paradigm with large-scale structures,}
  Phys.\ Rev.\ D {\bf 84} (2011) 063507
  [1103.2134].
  %%CITATION = ARXIV:1103.2134;%%

\bibitem{deSouza:2013wsa}
  R.~S.~de Souza,  {\it et al.},
%  A.~Mesinger, A.~Ferrara, Z.~Haiman, R.~Perna and N.~Yoshida,
  {\em Constraints on Warm Dark Matter models from high-redshift long gamma-ray bursts,}
  Mon.\ Not.\ Roy.\ Astron.\ Soc.\  {\bf 432} (2013) 3218
  [1303.5060].
  %%CITATION = ARXIV:1303.5060;%%

\bibitem{Mangano:2001iu}
  G.~Mangano, G.~Miele, S.~Pastor and M.~Peloso,
  {\em A Precision calculation of the effective number of cosmological neutrinos,}
  Phys.\ Lett.\ B {\bf 534} (2002) 8
  [astro-ph/0111408].
  %%CITATION = ASTRO-PH/0111408;%%

    \bibitem{Ade:2015xua}
  P.~A.~R.~Ade {\it et al.},
{\em Planck 2015 results. XIII. Cosmological parameters,}
  [1502.01589].
  %%CITATION = ARXIV:1502.01589;%%
  
\bibitem{Linde:2005ht}
  A.~D.~Linde,
{\em Particle physics and inflationary cosmology,}
  Contemp.\ Concepts Phys.\  {\bf 5} (1990) 1
  [hep-th/0503203].
  %%CITATION = HEP-TH/0503203;%%
  
\bibitem{B-via-L}  
%\bibitem{Kuzmin:1985mm}
  V.~A.~Kuzmin, V.~A.~Rubakov and M.~E.~Shaposhnikov,
   {\em On the Anomalous Electroweak Baryon Number Nonconservation in the Early Universe,}
  Phys.\ Lett.\ B {\bf 155} (1985) 36.
  %%CITATION = PHLTA,B155,36;%%
  %
%\bibitem{Fukugita:1986hr}
  M.~Fukugita and T.~Yanagida,
   {\em Baryogenesis Without Grand Unification,}
  Phys.\ Lett.\ B {\bf 174} (1986) 45.
  %%CITATION = PHLTA,B174,45;%%
  %
%\bibitem{Luty:1992un}
  M.~A.~Luty,
 {\em Baryogenesis via leptogenesis,}
  Phys.\ Rev.\ D {\bf 45} (1992) 455.
  %%CITATION = PHRVA,D45,455;%%
  
        \bibitem{Zurek:2013wia}
    For a review see, e.g~K.~M.~Zurek,
  {\em Asymmetric Dark Matter: Theories, Signatures, and Constraints,}
  Phys.\ Rept.\  {\bf 537} (2014) 91
  [1308.0338].
  %%CITATION = ARXIV:1308.0338;%%
  
  \bibitem{Abazajian:2013oma}
  K.~N.~Abazajian {\it et al.} 
  %[Topical Conveners: K.N. Abazajian, J.E. Carlstrom, A.T. Lee],
  {\em Neutrino Physics from the Cosmic Microwave Background and Large Scale Structure,}
  Astropart.\ Phys.\  {\bf 63} (2015) 66
  [1309.5383].
  %%CITATION = ARXIV:1309.5383;%%
  
    \bibitem{Barbieri:1999ma}
  R.~Barbieri, P.~Creminelli, A.~Strumia and N.~Tetradis,
  {\em Baryogenesis through leptogenesis,}
  Nucl.\ Phys.\ B {\bf 575} (2000) 61
  [hep-ph/9911315].
  %%CITATION = HEP-PH/9911315;%%
  
\bibitem{Buchmuller:2003gz}
  W.~Buchmuller, P.~Di Bari and M.~Plumacher,
{\em The Neutrino mass window for baryogenesis,}
  Nucl.\ Phys.\ B {\bf 665} (2003) 445
  [hep-ph/0302092].
  %%CITATION = HEP-PH/0302092;%%

\bibitem{Harvey:1990qw}
  J.~A.~Harvey and M.~S.~Turner,
{\em Cosmological baryon and lepton number in the presence of electroweak fermion number violation,}
  Phys.\ Rev.\ D {\bf 42} (1990) 3344.
  %%CITATION = PHRVA,D42,3344;%%
  
\bibitem{Holdom:1985ag}
  B.~Holdom,
  {\em Two U(1)'s and Epsilon Charge Shifts,}
  Phys.\ Lett.\ B {\bf 166} (1986) 196.
  %%CITATION = PHLTA,B166,196;%%

\bibitem{Kuzmin:1997jua}
  V.~A.~Kuzmin and V.~A.~Rubakov,
{\em Ultrahigh-energy cosmic rays: A Window to postinflationary reheating epoch of the universe?,}
  Phys.\ Atom.\ Nucl.\  {\bf 61} (1998) 1028
  [astro-ph/9709187].
  %%CITATION = ASTRO-PH/9709187;%%
 
\bibitem{Carone:2010ha}
  C.~D.~Carone, J.~Erlich and R.~Primulando,
{\em Decaying Dark Matter from Dark Instantons,}
  Phys.\ Rev.\ D {\bf 82} (2010) 055028
  [1008.0642].
  %%CITATION = ARXIV:1008.0642;%%

\bibitem{'tHooft:1976up}
  G.~'t Hooft,
{\em Symmetry Breaking Through Bell-Jackiw Anomalies,}
  Phys.\ Rev.\ Lett.\  {\bf 37} (1976) 8.
  %%CITATION = PRLTA,37,8;%%
    
  
  
  \end{thebibliography}
\end{document}